\documentclass[aps,superscriptaddress,prd,
onecolumn,
floatfix, 
nofootinbib,
amsmath,amssymb,amsfonts,longbibliography
]{revtex4-2}
\usepackage[pdftex]{graphicx}
\usepackage[colorlinks,linkcolor=blue,anchorcolor=violet,citecolor=red]{hyperref}

\usepackage{float,color}
\usepackage{rsfso}
\usepackage{physics}
\usepackage{comment}
\newcommand{\ep}{\epsilon}
\newcommand{\al}{\alpha}
\newcommand{\pa}{\partial} 
\newcommand{\del}{\delta}

\date{\today}
\begin{document}
\title{Large violation of Leggett--Garg inequalities 
with 
coherent-state projectors 
for a harmonic oscillator and chiral scalar field}
\author{Tomoya Hirotani}
\email{hirotani.tomoya.937@s.kyushu-u.ac.jp}
\author{Akira Matsumura}
\email{matsumura.akira@phys.kyushu-u.ac.jp}
\affiliation{Department of Physics,  Kyushu University, 744 Motooka, Nishi-Ku, Fukuoka 819-0395, Japan}
\author{Yasusada Nambu}
\email{nambu@gravity.phys.nagoya-u.ac.jp}
\affiliation{Department of Physics, Graduate School of Science, Nagoya
  University, Chikusa, Nagoya 464-8602, Japan}
\author{Kazuhiro Yamamoto}
\email{ yamamoto@phys.kyushu-u.ac.jp}
\affiliation{Department of Physics,  Kyushu University, 744 Motooka, Nishi-Ku, Fukuoka 819-0395, Japan}
\begin{abstract} 
We investigate violations of Leggett--Garg inequalities (LGIs) for 
a harmonic oscillator and a $(1+1)$-dimensional chiral scalar field with coherent-state projectors, which is equivalent to a heterodyne-type measurement scheme. 
For the harmonic oscillator, we found that the vacuum and thermal states violated the LGIs by evaluating the two-time quasi-probability distribution function. In particular, we demonstrate that the value of the two-time quasi-probability reaches $-0.123$ for a squeezed coherent-state projector, which is equivalent to $98\%$ of the L\"{u}ders bound corresponding to the maximal violation of the LGIs.
We also find a violation of the LGIs for the local mode of a quantum chiral scalar field by constructing a coherent-state projector similar to the harmonic oscillator case. 
In contrast to the harmonic oscillator, the periodicity in the time direction of the quasi-probability disappears, which is related to the existence of quantum entanglement between the local mode and its complementary degrees of freedom. 
\end{abstract}

\maketitle

\section{Introduction}
 How can we be sure that a system is quantum mechanical? Violation of macroscopic realism (MR) is a concept that characterizes quantum mechanics. In a classical system, the state of an object exists even before the measurement is made. However, in quantum mechanical systems, it is not necessarily true that an object is fixed in a particular state before observation.
 How can we check whether MR is broken?
 Leggett--Garg inequalities (LGIs) were proposed by Leggett and Garg in 1985 to test whether a macroscopic system follows MR \cite{PhysRevLett.54.857}.  According to Halliwell \cite{Halliwell_2019}, the necessary and sufficient condition for MR to be valid is the existence of 12 two-time LGIs and 4 three-time LGIs, for a total of 16 LGIs. In other words, if any of the 16 LGIs are violated, we can confirm that MR does not hold in that system. If MR does not hold, we can conclude that the system is quantum. Conversely, note that we cannot conclude that there is no quantumness in a system in which MR appears to hold; that is, all 16 LGIs hold true.

 In MR theory, for a dichotomic variable $Q(t)$ that takes definite values of $\pm 1$, we consider the sequential measurement of $Q$ at times $t_1$ and $t_2$. Then, the following inequalities hold:
\begin{equation}
 (1+s_1Q(t_1))(1+s_2Q(t_2))\ge 0,
\end{equation}
where $s_1=\pm1$ and $s_2=\pm1$. Following the MR framework, a joint probability distribution function exists for the measurement results, and the existence of such a joint probability implies that we can simply average the above formula and obtain  two-time LGIs:
\begin{equation}
 1+s_1\expval{ Q_1}+s_2\expval{Q_2}+s_1s_2\expval{Q_1Q_2}\ge0,
\end{equation}
where $Q_1:=Q(t_1)$, $Q_2:=Q(t_2)$, and $\langle\cdot\rangle$ denotes the ensemble averages of the measurement results. In quantum mechanics, the above expression is promoted to be positivity of the two-time quasi-probability defined by
\begin{equation}
 q_{s_1s_2}:=\frac{1}{4}\left(1+s_1\expval{\hat Q_1}+s_2\expval{\hat Q_2}+\frac{1}{2}s_1s_2\expval{\{\hat Q_1,\hat Q_2\}}\right)=\mathrm{Re}
 \mathrm{Tr}\left[\hat M_{s_2}(t_2)\hat M_{s_1}(t_1)\hat\rho_0\right],
 \label{eq:q12}
\end{equation}
where $\hat M_s=(1+s\hat Q)/2$ is the projector  (measurement operator) for the dichotomic operator $\hat Q$ and $\hat\rho_0$ is the quantum state of the target system. The two-time LGIs are equivalent to $q_{s_1s_2}\ge 0$ and the negativities of the quasi-probability indicate a violation of MR (violation of the LGIs).

In terms of the expected values of the dichotomic operator, the
 quasi-probability is
 \begin{align}
   q_{s_1s_2}
   &=\frac{1}{8}\expval{(1+s_1\hat Q_1+s_2\hat Q_2)^2-1}\geq-\frac{1}{8}.
 \end{align}
 This quantity is always positive for classical systems, and its negativity implies a violation of MR.
 The  minimum value $-1/8$ is called the L\"{u}ders bound \cite{Budroni2013}, which
 corresponds to the maximal violation of the LGIs. 
 To verify the LGIs experimentally, we obtain the expectation values $\expval{\hat Q_{1,2}}$ and $\expval{\hat Q_1\hat Q_2}$ from the experimental data and then check whether 
 the quasi-probability (\ref{eq:q12}) is negative.

The two-time probabilities for these sequential measurements are
\begin{equation}
 p_{12}(s_1,s_2):=\mathrm{Tr}\left[\hat M_{s_2}(t_2)\hat M_{s_1}(t_1)\hat\rho_0\hat M_{s_1}(t_1)\right],
\end{equation}
which is a manifestly positively valued quantity but does not satisfy the following no-signaling-in-time condition (NSIT) \cite{PhysRevA.87.052115,PhysRevA.91.062103,PhysRevLett.116.150401}:
\begin{equation}
 p_2(s_2):=\mathrm{Tr}\left[\hat M_{s_2}(t_2)\hat\rho_0\right]=\sum_{s_1}p_{12}(s_1,s_2).
\end{equation}
For a pure initial state $\rho_0=\ket{\psi_0}\bra{\psi_0}$, $p_{12}$ fails to satisfy the probability sum rule because of the interference between the pairs of states $
\ket{\psi_{+}}=\hat M_{+}(t_2)\hat M_{s_1}(t_1)\ket{\psi_0}$ and $\ket{\psi_{-}}=\hat M_{-}(t_2)\hat M_{s_1}(t_1)\ket{\psi_0}$. Although the quasi-probability formally satisfies the probability sum rule
\begin{equation}
 p_2(s_2)=\sum_{s_1}q_{s_1s_2},
\end{equation}
 the value of quasi-probability can become negative, which implies quantum mechanical behavior of the system.

As typical systems, LGIs are specifically applied to quantum-harmonic oscillators (QHOs) \cite{2018PhRvL.120u0402B,2022PhRvA.105b2221M,Mawby2022b,2022arXiv221110318D,2023arXiv231016471H}. Assuming that the dichotomic variable $\hat Q$ is in the form of a sign function indicates that the quasi-probability for the ground state of a QHO does not violate both the two-time LGIs and the three-time LGIs \cite{2022PhRvA.105b2221M}.  However, by making the dichotomic operator a more coarse-grained or nontrivial form \cite{2023arXiv231016471H}, even for the ground state, there are some cases in which the  LGIs are violated; we will confirm this behavior in this study. This implies that the features of the measurement operator must be carefully considered when verifying MR of the system. In previous studies, LGIs have been applied to various types of Gaussian states: coherent,  thermal, squeezed, squeezed coherent, and thermally squeezed coherent \cite{2018PhRvL.120u0402B,2022PhRvA.105b2221M,Mawby2022b,2022arXiv221110318D,2023arXiv231016471H}. With the settings of \cite{Mawby2022b}, the violation of the LGIs reached 84\% of  the  L\"{u}ders  bound \cite{Budroni2013}, which provided the maximal value of  the violation of the LGIs. Moreover, using a different 
scheme with temporal correlations (called no-signaling in time) has shown that violations of MR can be verified independently of the mass, momentum, and frequency of the QHO   \cite{2022arXiv221110318D}.
Experimental verification of LGIs has been reported in \cite{2014PhRvL.112s0402A,2019PhRvA.100d2325M,knee2016strict}, mainly using qubit systems.
Although there are few analyses of LGIs  applied to quantum fields, Tani et al. \cite{2023arXiv231102867T} investigated the LGIs for  $(3+1)$-dimensional quantum fields and $(1+1)$-dimensional chiral massless scalar fields, and they concluded that, for the vacuum state of the field, no violation of the LGIs occurred without adopting a nontrivial dichotomic variable $\hat Q$.

 In this study, we investigated the violation of LGIs for harmonic oscillator and chiral scalar field systems with coherent-state measurements. As  a dichotomic observable for the LGIs, we adopted one defined via a Gaussian projector (Gaussian POVM) \cite{Mawby2022b,2022arXiv221110318D}. We will see that the Gaussian projector results in a greater violation of the LGIs compared to non-Gaussian-type projectors ({e.g.,} $\hat Q=\mathrm{sgn}(\hat x)$ \cite{2022PhRvA.105b2221M,2023arXiv231016471H,2018PhRvL.120u0402B,Mawby2022b}). For the scalar field system, we introduced a local oscillator mode assigned to a specified spatial region from the scalar field using a window function.  This window function was selected to respect the symplectic structure of the local mode. Because the local mode is defined as an  oscillator embedded in the total quantum field system, its state becomes mixed and evolves in a nonunitary manner. Therefore, we do not have a Hamiltonian that determines the evolution of the local mode required to evaluate the projectors at specified times. To overcome this difficulty, we express the local mode operators using the creation and annihilation operators of the original quantum field and take the summation over each wavenumber mode to obtain the quasi-probability of the local mode.

The remainder of this paper is organized as follows: In Sec. II, we present the violation of the LGIs for a harmonic oscillator with a Gaussian projector, and we explain why the LGIs are violated in comparison with the classical motion of the oscillator.  In Sec. III, we consider the LGIs in a (1+1)-dimensional chiral scalar field and derive a formula for the quasi-probability of the spatial local modes of the scalar field. We found that the LGIs violated the suitable projector parameters. We also discuss the  LGI experiment using quantum Hall 
systems. Section IV is devoted to a summary and conclusions.
Throughout this paper, we adopt  the unit  $c=\hbar=1$.
\section{Violation of LGI for a harmonic oscillator system
}

\subsection{Harmonic oscillator system}
 Let us consider the quasi-probability in a harmonic oscillator system \cite{2018PhRvL.120u0402B,2022PhRvA.105b2221M,Mawby2022b,2022arXiv221110318D,2023arXiv231016471H} with the following  Hamiltonian:
\begin{equation}
  \hat H=\frac{\hat p^2}{2}+\frac{\Omega^2}{2}\hat
    q^2=
  \Omega\left(\hat a^\dag\hat a+\frac{1}{2}\right),
\end{equation}
where $\hat p$ and $\hat q$ are the dimensionless momentum and position operators, respectively, and $\Omega$ is the angular frequency of the harmonic oscillator.
The annihilation and creation operators are defined as
\begin{equation}
  \hat a=\sqrt{\frac{\Omega}{2}}\,\hat q+\frac{i}{\sqrt{2\Omega}}\hat
  p,\quad
  \hat a^\dag=\sqrt{\frac{\Omega}{2}}\,\hat q-\frac{i}{\sqrt{2\Omega}}\hat p.
\end{equation}
The oscillator ground state is defined as $\hat
a\ket{0}=0$. For this ground state, each expectation value is calculated as $\expval{\hat q}=\expval{\hat p}=0,~\expval{\hat q^2}=1/(2\Omega)$, and $~\expval{\hat p^2}=\Omega/2$.
The coherent state of the oscillator is defined as
\begin{equation}
  \ket{\al}=e^{\al\hat a^\dag-\al^*\hat a}\ket{0}=\hat D_a(\al)\ket{0}=e^{-|\al|^2/2}\sum_n\frac{\al^n}{\sqrt{n!}}\ket{n},
\end{equation}
where $\hat D_a(\alpha)$ denotes the displacement operator. With the evolution
operator $\hat U(t)=\exp(-i\hat Ht)$, the coherent state evolves into
\begin{equation}
  \hat U(t)\ket{\al}=e^{-i\Omega t/2}\ket{e^{-i\Omega t}\al},\quad \hat U^\dag\hat
  a\hat U=e^{-i\Omega t}\hat a.
\end{equation}


For the initial state of the oscillator, we assume a thermally coherent
state
\begin{align}
  &\hat\rho_0=C \hat D_a(\al)\exp\left(-A \hat a^\dag\hat
    a\right)\hat D_a^\dag(\al),\quad\mathrm{Tr}\hat\rho_0=1,
    \label{eq:thco}
\end{align}
with the normalization  $C=\left(\mathrm{Tr}\exp\left(-A \hat a^\dag\hat
    a\right)\right)^{-1}=(1-e^{-A})$. The constant $A$ is related to the symplectic eigenvalue $\nu$ of the covariance matrix for the Gaussian state \eqref{eq:thco} as follows:
\begin{equation}
  e^{-A}=\frac{\nu-1/2}{\nu+1/2},\quad \nu^2=\expval{\hat
    q^2}\expval{\hat p^2}\ge \frac{1}{4}.
\end{equation}
The temperature $T$ of the thermal state is related to $\nu$ by $\nu^{-1}=2\tanh\left(\Omega/(2T)\right)$.
\footnote{We can use the following formula to express the thermal state in terms
of the displacement operator:
$$
  e^{-A\hat a^\dag\hat
    a}=
\frac{1}{(\nu+1/2)\pi}\int d^2z\,e^{-\nu|z|^2}\hat D_a(z),\quad
  \nu=\frac{1}{2}\left(\frac{1+e^{-A}}{1-e^{-A}}\right).
$$
}
In the zero-temperature limit, $T\rightarrow 0$, $\nu=1/2$ and $\hat\rho_0$ becomes a pure state.

\begin{figure}[b]
  \centering
  \includegraphics[width=0.7\linewidth]{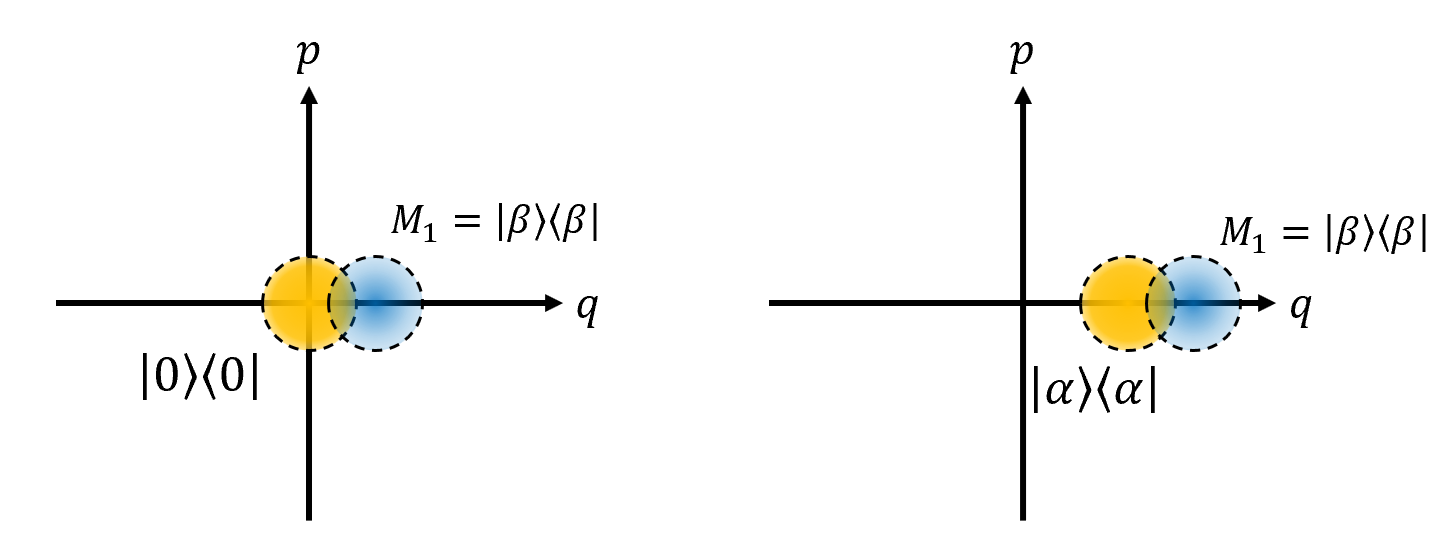}
  \caption{Schematic plot of the coherent-state measurement of the
    vacuum state $\ket{0}$ (left panel) and the coherent state $\ket{\alpha}$ (right panel).  The dashed circles represent $1\sigma$ contours of the Gaussian state. As the measurement result, we assign the dichotomic variable $Q=+1$ if the obtained measurement result
     $(q,p)$ is included in the blue disks; otherwise, $Q=-1$. After the measurement, the state $\ket{\al}\bra{\al}$ of the target system is projected to $|\bra{\al}\ket{\beta}|^2\ket{\beta}\bra{\beta}$ if the measurement result is $Q=+1$.}
  \label{fig:measurement}
 \end{figure}
\subsection{Measurement operator}

As discussed extensively in \cite{PhysRevA.93.022123}, violation of conventional LGIs is equivalent to the existence of negative values of quasi-probability defined by 
\begin{equation}
  q_{s_1s_2}=\mathrm{Re}\mathrm{Tr}\left[\hat M_{s_2}(t_2)\hat
    M_{s_1}(t_1)\hat
    \rho_0\right],
\end{equation}
where $\hat M_s$ is the measurement operator for a dichotomic observable
$\hat Q$ with the specified measurement results $s=\pm 1$ and $\hat\rho_0$
denotes the initial state of the target oscillator. In terms of $\hat Q$,
the measurement operator is expressed as
\begin{equation}
  \hat M_s=\frac{1+s\,\hat Q}{2},\quad \hat Q^2=1,\quad s=\pm1.
\end{equation}
We consider LGIs with generalized measurements by adopting the following Gaussian measurement
operators \cite{Adesso2014}
\begin{equation}
  \hat M_s=\frac{1-s}{2}+s\ket{\beta_b}\bra{\beta_b}=
  \begin{cases}
    \ket{\beta_b}\bra{\beta_b}&\quad \text{for}\quad s=+1,\\
    1-\ket{\beta_b}\bra{\beta_b}&\quad \text{for}\quad s=-1,
  \end{cases}
  \label{eq:projector}
\end{equation}
where $\ket{\beta_b}$ is the coherent state with the annihilation operator defined by
\begin{equation}
  \hat b:=\sqrt{\frac{\omega}{2}}\,\hat q+\frac{i}{\sqrt{2\omega}}\hat
  p=\cosh r\,\hat a+\sinh r\,\hat a^\dag, \quad e^r=\sqrt{\frac{\omega}{\Omega}},
  \label{eq:QHOsqp}
\end{equation}
where $\omega$ is the angular frequency of the projector (a measurement apparatus parameter). This type of measurement operator maps Gaussian states to Gaussian states and can be realized experimentally by appending ancillary modes initialized in Gaussian states, implementing Gaussian unitary operations on the system and ancillary modes, and measuring quadrature operators that can be achieved using balanced homodyne detection in the optics framework. The
annihilation operator is expressed as $\hat b=\hat S^\dag(r)\,\hat a\,\hat
  S(r)$ with  the squeezing operator
\begin{equation}
  \hat S(\zeta)=\exp\left(\frac{\zeta}{2}\hat
    a^\dag{}^2-\frac{\zeta^*}{2}\hat a^2\right),\quad \zeta=re^{i\varphi},
\end{equation}
where $\zeta$ is the squeezing parameter and $\varphi$ is the constant phase.
The ``vacuum state'' defined by $\hat b$ is determined as $\hat
b\ket{0_b}=0$ with $\ket{0_b}=\hat S^{-1}(r)\ket{0}$, where $\hat a\ket{0}=0$. Hence
\begin{equation}
  \ket{\beta_b}=e^{\beta\hat b^\dag-\beta^*\hat b}\hat
  S(-r)\ket{0}=\hat D_a(\gamma)\hat
  S(e^{i\pi}r)\ket{0}=:\ket{\gamma,\zeta},\quad \gamma=\beta\cosh
  r-\beta^*\sinh r,\quad \zeta=e^{i\pi}r.
\end{equation}
Therefore, $\ket{\beta_b}$ is the squeezed coherent state
$\ket{\gamma,\zeta}$ with displacement parameter $\gamma$ and squeezing parameter $\zeta$. Parameters $\beta$, $r$, and $\omega$ in the projector \eqref{eq:projector}
are regarded as the measurement parameters. 
The evolution of the projector $\hat M_1=\ket{\beta_b}\bra{\beta_b}$ is
\begin{align}
  \hat M_1(t)
  &=\hat U^\dag(t)\hat M_1\hat U(t) \notag \\
  &=\hat U^\dag(t)\hat D(\gamma)\hat S(\zeta)\ket{0}\bra{0}\hat
    S^\dag(\zeta)\hat D^\dag(\gamma)\hat U(t) \notag \\
  &=\hat D(e^{i\Omega t}\gamma)\hat S(e^{2i\Omega t}\zeta)\ket{0}\bra{0}\hat
    S^\dag(e^{2i\Omega t}\zeta)\hat D^\dag(e^{i\Omega t}\gamma)
    \notag \\
  &=\ket{e^{i\Omega t}\gamma, e^{2i\Omega t}\zeta}\bra{e^{i\Omega t}\gamma, e^{2i\Omega t}\zeta}.
\end{align}
The projector with measurement result $s$ is
\begin{equation}
  \hat
  M_s(t)=\frac{1-s}{2}+s\ket{\gamma(t),\zeta(t)}\bra{\gamma(t),\zeta(t)},\quad
  \gamma(t)=e^{i\Omega t}\gamma,\quad \zeta(t)=e^{2i\Omega t+i\pi}r.
  \label{eq:mop}
\end{equation}
\begin{figure}[t]
  \centering
  \includegraphics[width=0.9\linewidth]{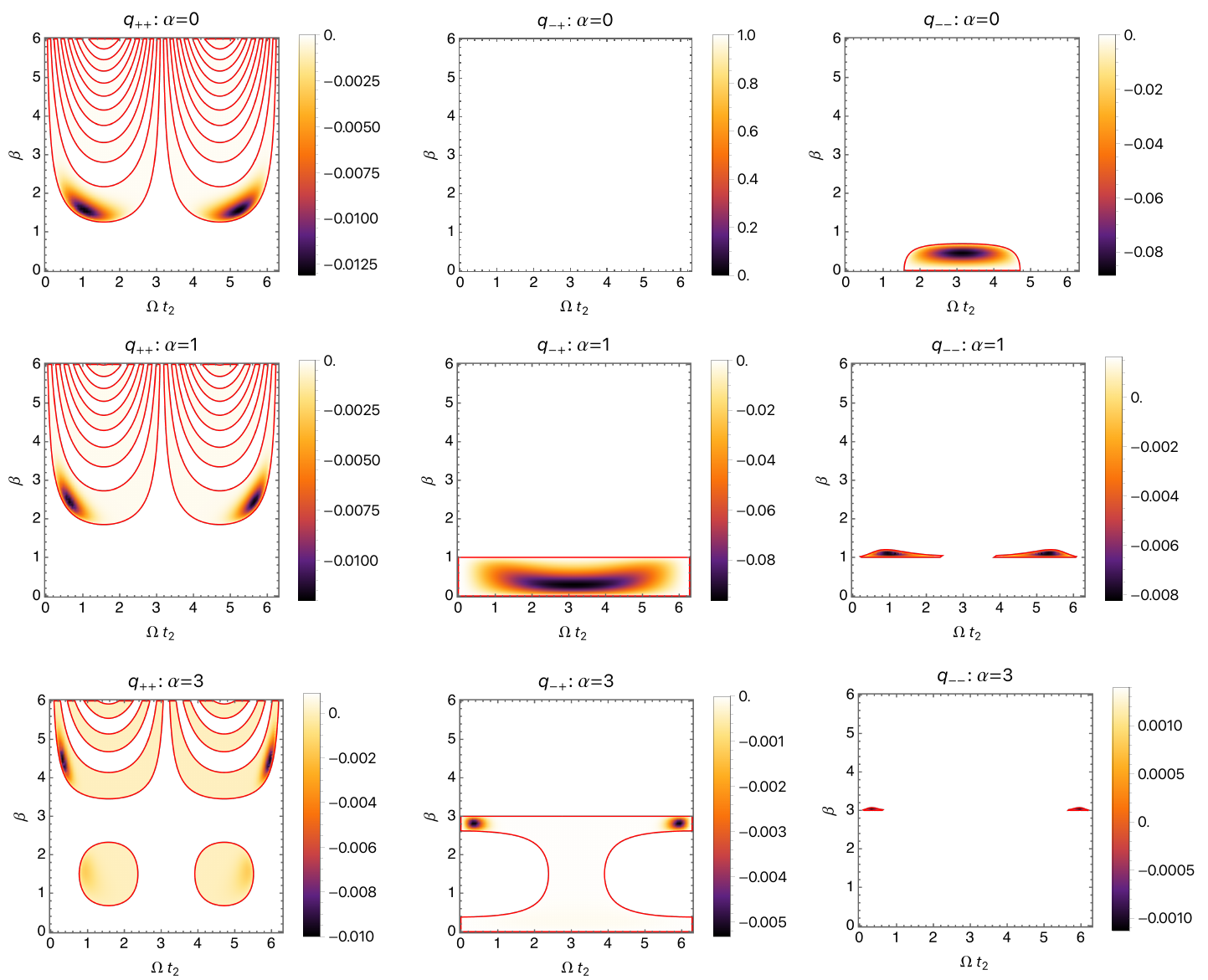}
  \caption{Negative regions  of the quasi-probability (enclosed by red lines) for  the pure  state $\ket{\alpha}$ with $\alpha=0,$ 1, and 3.    For $\alpha=0$, $q_{++}$ and
    $q_{--}$ become negative. For $\alpha\neq 0$, $q_{++},$ $q_{--},$ and $q_{+-}$ become negative. We chose $t_1=0$ and  $\omega=\Omega$.}
  \label{fig:LGI2}
\end{figure}
If we adopt the dichotomic operator $\hat Q=2\ket{\beta}\bra{\beta}-1$, where $\ket{\beta}$ is a coherent state, then the measurement operator for this dichotomic operator is $\hat M_1=\ket{\beta}\bra{\beta}$ and $\hat
M_{-1}=1-\ket{\beta}\bra{\beta}$
(Fig. \ref{fig:measurement}). Hence for the vacuum state of the target oscillator, if the measurement result with a fixed
$s$ is $+1$, this implies $Q=+1$; on the other hand, if the
measurement result is $-1$, this implies $Q=-1$. Physical interpretation
of this measurement is as follows: If the  measured values $(q,p)$  of the oscillator (yellow-shaded disks in Fig. \ref{fig:measurement}) are included in the width of the coherent-state  projector (blue-shaded disks in Fig. \ref{fig:measurement}), we will obtain the
measurement result $+1$; otherwise, the measurement result is $-1$.

\subsection{Quasi-probability with a coherent-state projector for the initial thermal coherent state
}
By using the measurement operator \eqref{eq:mop}, the quasi-probability
with an initial thermal coherent state \eqref{eq:thco} becomes 
\begin{align}
  q_{s_1s_2}&=C\,\mathrm{Re}\mathrm{Tr}\left[\hat M_{s_2}(t_2)\hat
              M_{s_1}(t_1)\hat D(\al)e^{-A\hat a^\dag\hat
              a}\hat D^\dag(\al)\right] \notag \\
  &=C\,\mathrm{Re}\mathrm{Tr}\left[\hat D^\dag(\al)\hat M_{s_2}(t_2)\hat
    D(\al)\,\hat D^\dag(\al)\hat M_{s_1}(t_1)\hat D(\al)e^{-A\hat
    a^\dag\hat a}\right],
    \label{eq:q-general0}
 \end{align}
where
\begin{equation}
  \hat D^\dag(\al)\hat M_s(t)\hat D(\al)
  =\frac{1-s}{2}+s\ket{\gamma(t)-\al,\zeta(t)}\bra{\gamma(t)-\al,\zeta(t)}.
\end{equation}
Then, using $ e^{-A\hat a^\dag\hat a}=\sum_n
  e^{-A\,n}\ket{n}\bra{n}$  in Eq. \eqref{eq:q-general0} gives
\begin{align}
  q_{s_1s_2}
  &=C\,\mathrm{Re}\sum_n\Bigl[\frac{(1-s_1)(1-s_2)}{4}e^{-A\,n}
              +\frac{s_1(1-s_2)}{2}e^{-A\,n}
              |\langle n|\gamma_1-\al,\zeta_1\rangle|^2
              \nonumber\\
              &\qquad\qquad\qquad +\frac{s_2(1-s_1)}{2}e^{-A\,n}
              |\langle n|\gamma_2-\al,\zeta_2\rangle|^2 \notag \\
 &\qquad\qquad\qquad +s_1s_2
   e^{-A\,n}\bra{\gamma_2-\al,\zeta_2}\ket{\gamma_1-\al,\zeta_1}\bra{n}
   \ket{\gamma_2-\al,\zeta_2}\bra{\gamma_1-\al,\zeta_1}\ket{n}\Bigr].
\end{align}
For a coherent-state projector without squeezing, we set
$\zeta=0$ and $ \gamma=\beta$ and the quasi-probability is evaluated as
\begin{align}
 q_{s_1s_2} &=C\,\mathrm{Re}\Bigl[\frac{(1-s_1)(1-s_2)}{4}\sum_n e^{-A\,n}
    \notag \\
  &\qquad\qquad
    +\frac{s_1(1-s_2)}{2}e^{-|\beta_1-\al|^2}\sum_ne^{-A\,n}\frac{|\beta_1-\al|^{2n}}{n!}
    +\frac{s_2(1-s_1)}{2}e^{-|\beta_2-\al|^2}\sum_ne^{-A\,n}\frac{|\beta_2-\al|^{2n}}{n!}
    \notag \\
            &\qquad\qquad
              +s_1s_2\bra{\beta_2-\al}\ket{\beta_1-\al}e^{-|\beta_2-\al|^2/2-|\beta_1-\al|^2/2}
              \sum_ne^{-A\,n}\frac{(\beta_2-\al)^n(\beta_1^*-\al^*)^n}{n!}\Bigr]
              \notag\\
  &=C\,\mathrm{Re}\Bigl[\frac{(1-s_1)(1-s_2)}{4}\frac{1}{1-e^{-A}}
    \notag \\
  &\qquad\qquad
    +\frac{s_1(1-s_2)}{2}e^{-|\beta_1-\al|^2}\exp\left(e^{-A}\,|\beta_1-\al|^2\right)
    +\frac{s_2(1-s_1)}{2}e^{-|\beta_2-\al|^2}\exp\left(e^{-A}|\beta_2-\al|^2\right)
    \notag \\
  &\qquad\qquad
              +s_1s_2e^{-|\beta_2-\al|^2-|\beta_1-\al|^2}e^{(\beta_2^*-\al^*)(\beta_1-\al)}
             \exp\left(e^{-A}(\beta_2-\al)(\beta_1^*-\al^*)\right)
    \Bigr] \notag \\
   &=\frac{(1-s_1)(1-s_2)}{4}
    +(1-e^{-A})\Bigl[\frac{s_1(1-s_2)}{2}e^{-|\beta_1-\al|^2}\exp\left(e^{-A}\,|\beta_1-\al|^2\right)
    \notag \\
  &\qquad\qquad
    +\frac{s_2(1-s_1)}{2}e^{-|\beta_2-\al|^2}\exp\left(e^{-A}|\beta_2-\al|^2\right)
    \notag \\
  &\qquad\qquad
              +s_1s_2e^{-|\beta_2-\al|^2-|\beta_1-\al|^2}\mathrm{Re}\,e^{(\beta_2^*-\al^*)(\beta_1-\al)}
             \exp\left(e^{-A}(\beta_2-\al)(\beta_1^*-\al^*)\right) \Bigr],
\end{align}
where $\beta_1:=e^{i\Omega t_1}\beta$ and $\beta_2:=e^{i\Omega t_2}\beta$.
For the zero-temperature limit $e^{-A}\rightarrow 0$,
\begin{align}
q_{s_1s_2}
  &=\frac{(1-s_1)(1-s_2)}{4}+\frac{s_1(1-s_2)}{2}e^{-|\beta_1-\al|^2}
    +\frac{s_2(1-s_1)}{2}e^{-|\beta_2-\al|^2}
    \notag \\
  &\qquad
    +s_1s_2e^{-|\beta_2-\al|^2-|\beta_1-\al|^2}\,
    \mathrm{Re}\,e^{(\beta_2^*-\al^*)(\beta_1-\al)}.
    \label{eq:q-coherent}
\end{align}
Furthermore, for the vacuum state $\alpha=0$, the quasi-probability $q_{s_1s_2}$ depends only on the difference between the two measurement times $\Omega(t_2-t_1)$ and $\beta$.
For the high-temperature limit $e^{-A}\rightarrow 1$, the quasi-probability becomes $q_{s_1s_2}=(1-s_1)(1-s_2)/4\ge 0$ and the LGIs are not violated within this limit.
\begin{figure}[t]
  \centering
  \includegraphics[width=0.6\linewidth]{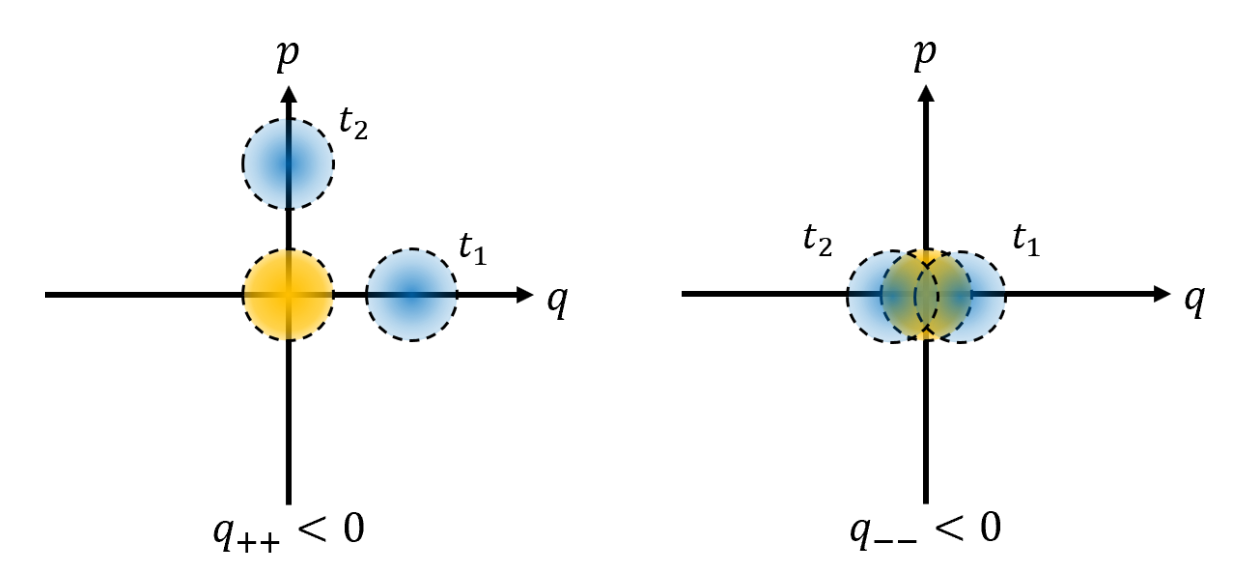}
  \caption{Schematic explanation of violation of the LGIs for the coherent-state projector. Yellow regions represent the initial state of the oscillator, and blue regions represent $1\sigma$ contours of the projector. The left panel represents the projector with $\beta>1$ and the right panel represents the projector with $\beta<1$.}
  \label{fig:LLGI2}
\end{figure}

The features and interpretation of violation of the LGIs for the initial vacuum state, initial coherent state, and initial thermal state are as follows: 
\paragraph{Initial vacuum state $\al=0$:}
The first row in Fig. \ref{fig:LGI2}  shows the behaviour of  the
quasi-probability with an initial vacuum state $\ket{0}$
(Eq. \eqref{eq:q-coherent}), where we chose $t_1=0$ and $\omega=\Omega$.  Regions
where $q_{s_1s_2}<0$ (enclosed by red lines) is in the $(t,\beta)$ plane. Figure \ref{fig:LLGI2} shows a schematic interpretation of the violation of the LGIs with a coherent projector.
In this case, $q_{++}$ and $q_{--}$ are negative. For a projector with parameter $\beta<1$, because the target vacuum state of the oscillator is localized at $q\sim 0$, the measurement discriminates whether the particle is located around $q\sim 0$ (measurement with $s=+1$) or not (measurement result with $s=-1$). Hence, $q_{++}$ does not violate the LGIs with this value of the parameter $\beta$; Immediately after the first measurement, the particle is definitely located around $q\sim 0$ and the second measurement will result in the same result as the first measurement, and these measurement results are classically expected. However, $q_{--}$ represents the ``possibility" that a particle does not exist around $q=0$ for both the first and second measurements, and such a measurement result cannot be expected classically. Therefore, $q_{--}$ becomes negative and MR is violated. Furthermore, it can be expected that the maximum violation occurs at the half period $\Omega\, t_2=\pi$ because of the largest overlap between the yellow region (state of the oscillator) and blue region (projector) in the right panel of Fig. \ref{fig:LLGI2}.
For the $\beta=0$ case, $q_{s_1s_2}=(1+s_1)(1+s_2)/4\geq0$ and there is no violation of the LGIs. This is because the measurement with $\beta=0$ does  not disturb the target vacuum state, and the obtained measurement results coincide with the classical results (and the particle always exists at $q=0$ in this case). 
For $\beta>1$, $q_{++}$ is negative. Negative regions in $q_{++}$ form a fringe structure that implies  quantum interference of the oscillator motion; indeed, the quasi-probability for the initial vacuum state is regarded as overlap of two branches of evolved states $\ket{\psi_1}:=\hat M_{+}(t_1)\ket{0}$ and $\ket{\psi_2}:=\hat M_{+}(t_2)\ket{0}$, and $\bra{\psi_1}\ket{\psi_2}$ represents interference of these two branches. After the first measurement at $t_1$, the center of the wavefunction is projected to the location where $\beta>1$ and then evolves according to the classical equation of motion. Because classical particle motion is  oscillatory in time,  the measurement result $Q_{s_1=1}=Q_{s_2=1}=+1$  cannot  be expected classically. The measurement result depends on the value of $t_2-t_1$. Therefore, $q_{++}$ was negative. 

\begin{figure}[H]
  \centering
  \includegraphics[width=0.7\linewidth]{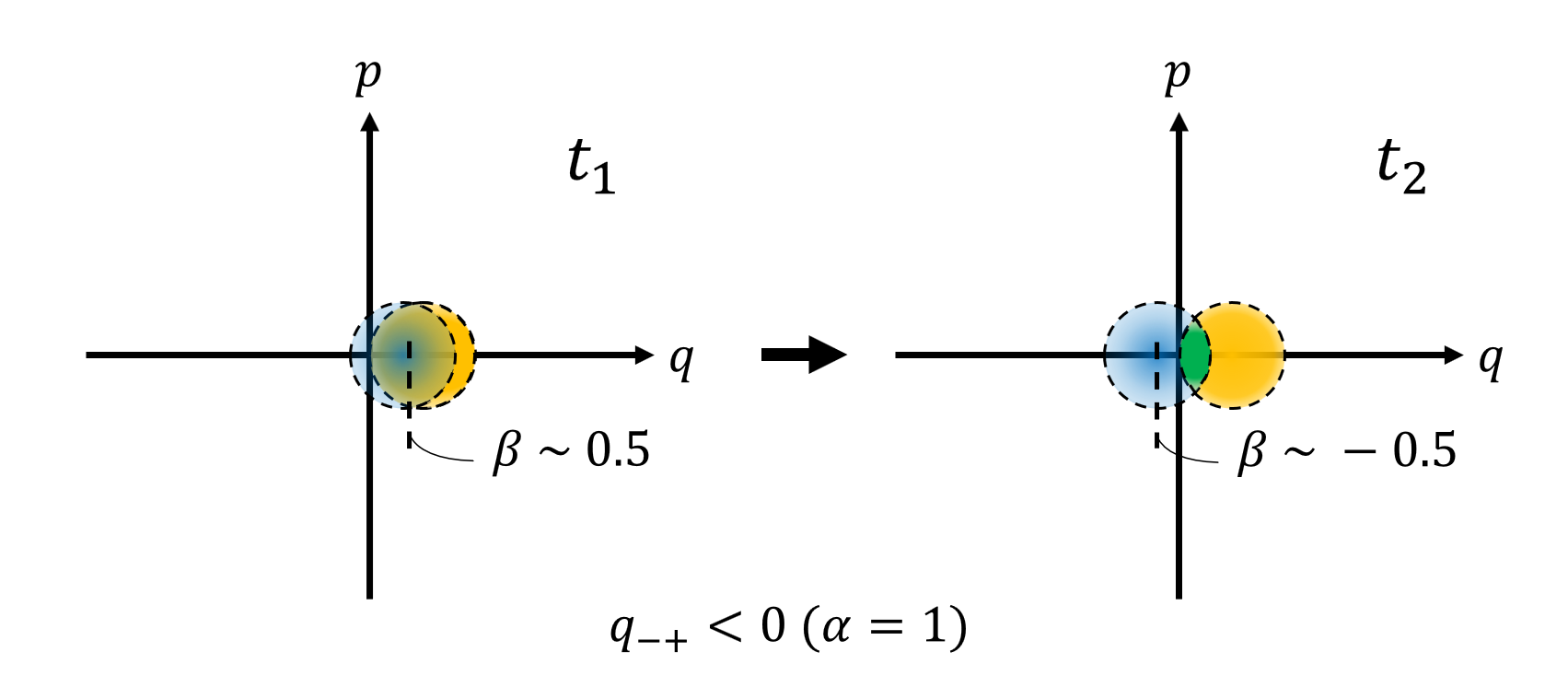}
  \caption{Schematic explanation of violation of the LGIs for  $q_{-+}$ with $\alpha = 1$ and $\beta \sim 0.5$. At time $t=t_1$, $s_1=-1$, and the particle is expected to be in the yellow region. Then, at time $t=t_2$, the particle is expected to be in the green region because $s_2=+1$. The mismatch in location in the phase space between the yellow region on the left panel $(t=t_1)$ and the green region on the right panel $(t=t_2)$ leads to the violation of the LGIs.}
  \label{fig:LGIq-+}
\end{figure}

\paragraph{Initial coherent state $\al\neq 0$:} 
The second and third  rows in Fig. \ref{fig:LGI2}  show the behaviour of  the
quasi-probability with an initial coherent state $\ket{\alpha}$
(Eq. \eqref{eq:q-coherent}), where we chose $t_1=0$ and $\omega=\Omega$.  Regions
where $q_{s_1s_2}<0$ (enclosed by red lines) are in the $(t,\beta)$ plane. In this case, $q_{++}$, $q_{-+}$, and $q_{--}$ are negative. We can provide the following explanation for the emergence of negative regions in the quasi-probability: $q_{++}$ exhibits a fringe pattern in the region $\beta>\alpha$, which is the same behavior as that in the $\alpha=0$ case. 
Because the quasi-probability with the initial coherent state overlaps the two states $\hat M(t_1)\ket{\alpha}$ and $\hat M(t_2)\ket{\alpha}$, it represents interference between these two states. 
For $\beta>\alpha$, the interference fringe appears as a negative value for $q_{++}$ in the $(t,\beta)$ plane. If we focus only on the position of the oscillator,  where the value of the violation is relatively large (dark black region in Fig. \ref{fig:LGI2}), then we observe that the relation $\beta \sim \alpha \pm 1.5 ~(>1)$  holds with any $\alpha$. This relationship  explains why round regions with a negative quasi-probability  suddenly appear in the lower left panel of Fig. \ref{fig:LGI2}. Although not included in the figure, as can be seen from the relation   $\beta\sim\alpha\pm1.5$, round regions with negative quasi-probabilities begin to appear when $\alpha$ begins to exceed $\sim$$2.5$. Physically, one could say that an overlap between the state of the oscillator and the projector that is neither too large nor too small  is required for a violation of the LGIs (Fig. \ref{fig:LLGI2}). 
For $q_{-+}$, if we select the $\alpha=1$ case as shown in Fig.  \ref{fig:LGIq-+}, the first measurement result implies that the particle exists around the yellow region ($Q=-1$), but the second measurement confirms that the particle exists around the green region ($Q=+1$), which is not classically expected. Of course, we must consider the spread of the oscillator's wavefunction; this effect is omitted here for intuitive understanding.
The quasi-probability $q_{--}$ can become negative in narrow regions with  $\alpha \sim \beta$ in the $(t,\beta)$ plane. That the first and second measurement results predict the nonexistence of the particle around $q=0$ is not classically expected because the yellow region (state of the oscillator) and blue regions (projector) overlap  (Fig. \ref{fig:LLGI2}). Therefore, $q_{--}$ has negative values and the LIGs are violated.

\begin{figure}[H]
  \centering
  \includegraphics[width=0.80\linewidth]{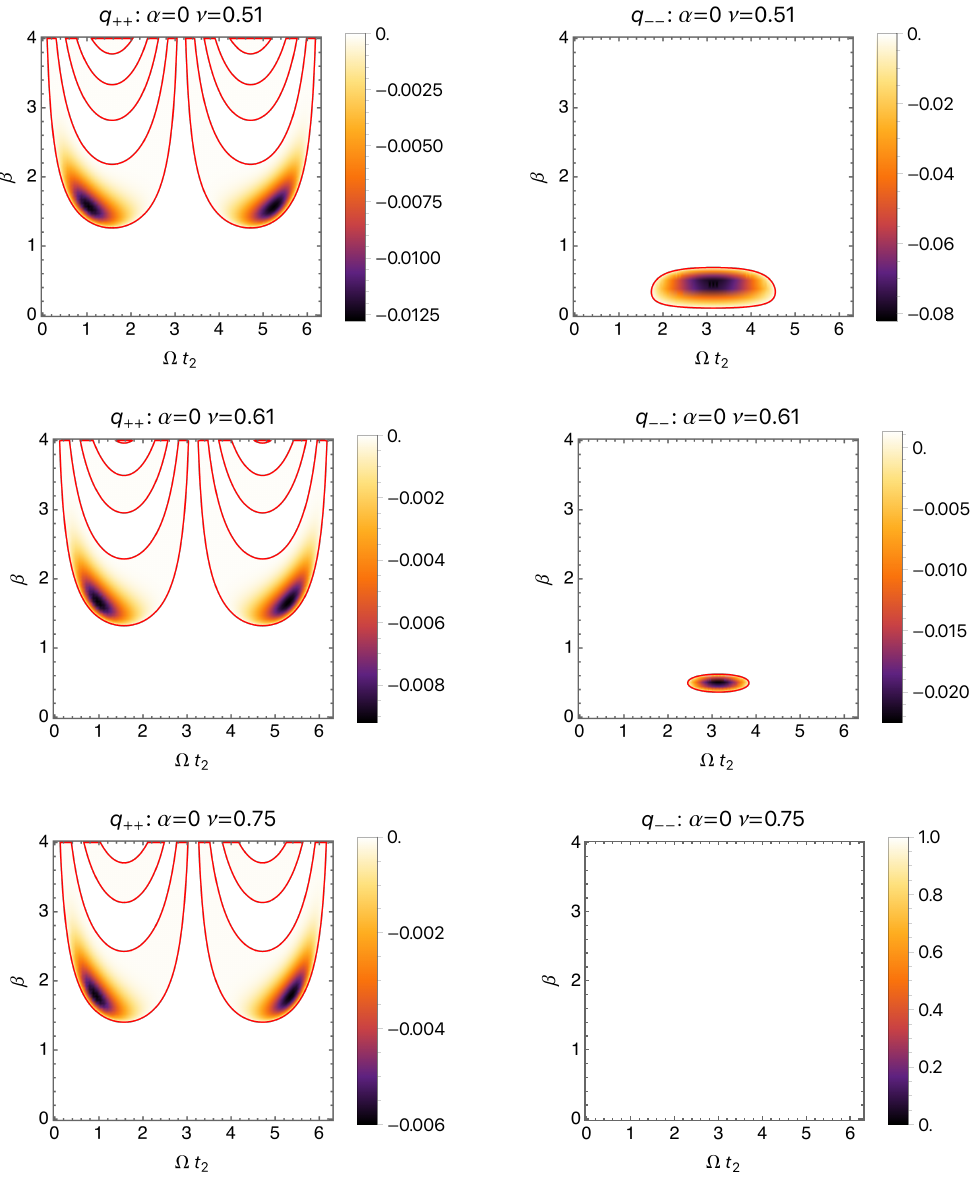}
  \caption{Negative regions of the quasi-probability (enclosed by red lines) for the initial
    thermal state $\alpha=0$  with  $\nu=0.51,$ 0.61, and 0.75. $\nu=0.5$ corresponds to a pure state. As $\nu$ increases, the thermal effect
    reduces the violation of the LGIs. We chose $t_1=0$ and $\omega=\Omega$.}
  \label{fig:qt1}
\end{figure}

\begin{figure}[H]
  \centering
  \includegraphics[width=0.85\linewidth]{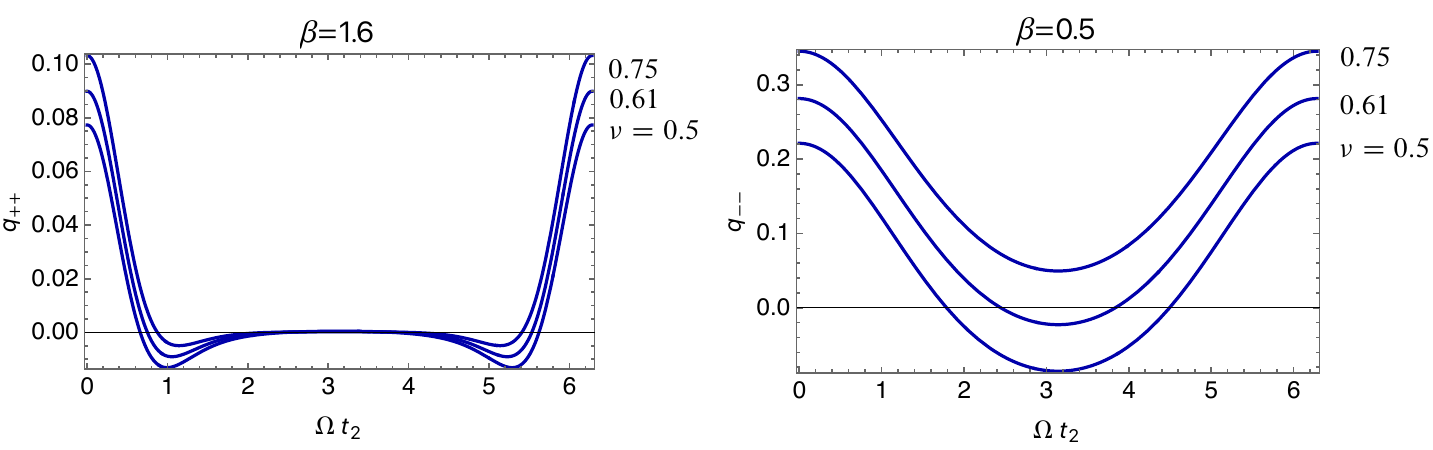}
  \caption{Time dependence of the quasi-probability $q_{++}$ with $\beta=1.6$ (left panel) and $q_{--}$ with $\beta=0.5$ (right panel) for the initial
    thermal state with $\nu=0.5,$ 0.61,and 0.75. $\nu=0.5$ corresponds to a pure state. The thermal effect
    reduces violation of the LGIs. For $\nu=0.75$, $q_{--}$ does not have  negative values.}
  \label{fig:qt2}
\end{figure}

\paragraph{Initial thermal state:}
Figure \ref{fig:qt1} shows the thermal effect on the negativity of the quasi-probability. For the initial mixed states with $\nu=0.51$, 0.61, and 0.75,  the negative regions of $q_{s_1s_2}$ become smaller with an increase in  $\nu$ compared with those of the initial vacuum state. Therefore, the  thermal effect (mixedness of the initial states)
reduces violation of the LGIs. Above the critical value of $\nu$, violation of the 
LGIs is not observed in $q_{--}$. Figure \ref{fig:qt2} shows the time dependence of $q_{++}$ (with $\beta=1.6$) and $q_{--}$ (with $\beta=0.5$) for $\nu=0.5,$ 0.61, and 0.75. This shows that the smallest negative values of the quasi-probability approach zero with an increase in $\nu$ and the thermal effect prevents violation of the LGIs.

\subsection{Quasi-probability with a squeezed coherent-state projector
  in the initial vacuum state}
Now, we consider a projector with a squeezed coherent state \eqref{eq:mop} for the initial vacuum state.  After taking the zero-temperature limit $A\rightarrow\infty$ and
$\al=0$ in Eq. \eqref{eq:q-general0}, the quasi-probability with the
initial vacuum state is
\begin{align}
  q_{s_1s_2}
  &=\mathrm{Re}\Bigl[\frac{(1-s_1)(1-s_2)}{4}
              +\frac{s_1(1-s_2)}{2}
              |\bra{0}\ket{\gamma_1,\zeta_1}|^2+\frac{s_2(1-s_1)}{2}
              |\bra{0}\ket{\gamma_2,\zeta_2}|^2 \notag \\
  &\qquad\qquad\qquad +s_1s_2
    \bra{\gamma_2,\zeta_2}\ket{\gamma_1,\zeta_1}\bra{0}\ket{\gamma_2,\zeta_2}\bra{\gamma_1,\zeta_1}
    \ket{0}\Bigr]
    ,
 \end{align}
 where\footnote{We use the following formulas for the squeezed coherent state 
   $\ket{\beta,\xi}$, where $\xi=e^{i\varphi}r$ \cite{Mo/ller1996}:
   \begin{align*}
     &\bra{0}\ket{\beta,\xi}=\frac{1}{\sqrt{\cosh
       r}}\exp\left[-\frac{1}{2}|\beta|^2+\frac{e^{i\varphi}}{2}\beta^*{}^2\tanh
       r\right], \\
     &\bra{\beta_1,\xi_1}\ket{\beta_2,\xi_2}=\frac{1}{\sqrt{\sigma_{21}}}\exp\left[
     \frac{\eta_{21}\eta_{12}^*}{2\sigma_{21}}+\frac{1}{2}\left(\beta_2\beta_1^*-\beta_2^*\beta_1\right)
       \right], 
   \end{align*}
 for $ r_2=r_1\equiv r$,   where
   \begin{align*}
     &\sigma_{21}=\cosh^2r-e^{i(\varphi_2-\varphi_1)}\sinh^2r,\\
     &\eta_{21}=(\beta_2-\beta_1)\cosh
       r-(\beta_2^*-\beta_1^*)e^{i\varphi_2}\sinh r,\quad
       \eta_{12}=(\beta_1-\beta_2)\cosh
       r-(\beta_1^*-\beta_2^*)e^{i\varphi_1}\sinh r
   \end{align*}
 }
 \begin{align}
   &\bra{0}\ket{\gamma_j,\zeta_j}=\frac{1}{\sqrt{\cosh
     r}}\exp\left[-\frac{\gamma^2}{2}(1+\tanh r)\right].
 \end{align}
 has no time dependence. Therefore, the quasi-probability is given by
 \begin{align}
   q_{s_1s_2}
   &=\frac{(1-s_1)(1-s_2)}{4}+|\bra{0}\ket{\gamma,\zeta}|^2\left[
     \frac{s_1(1-s_2)}{2}+\frac{s_2(1-s_1)}{2}+s_1s_2\mathrm{Re}\bra{\gamma_2,\zeta_2}
     \ket{\gamma_1,\zeta_1}\right],
     \label{eq:qsc}
 \end{align}
 with
 \begin{align}
   &\bra{\gamma_2,\zeta_2}
   \ket{\gamma_1,\zeta_1}
  =\frac{1}{\sqrt{\sigma_{21}}}\exp\left[
    \frac{\eta_{21}\eta_{12}^*}{2\sigma_{21}}+\frac{1}{2}\left(\gamma_2\gamma_1^*
    -\gamma_2^*\gamma_1\right) 
    \right], \\
   &\qquad \sigma_{21}=\cosh^2r-e^{2i\Omega(t_2-t_1)}\sinh^2r,\\
   &\qquad \eta_{21}=(e^{i\Omega t_2}-e^{i\Omega t_1})\gamma \cosh
     r-(e^{-i\Omega t_2}-e^{-i\Omega t_1})\gamma e^{i(2\Omega
     t_2+\pi)}\sinh r,\\
   &\qquad
     \eta_{12}=-(e^{i\Omega t_2}-e^{i\Omega t_1})\gamma \cosh
     r+(e^{-i\Omega t_2}-e^{-i\Omega t_1})\gamma e^{i(2\Omega
     t_1+\pi)}\sinh r,\\
   &\qquad
     \gamma_2\gamma_1^*=e^{i\Omega(t_2-t_1)}\gamma^2,
 \end{align}

 \begin{align}
   &\eta_{21}\eta_{12}^*=\gamma^2e^{-i\Omega(t_2-t_1)}(1-e^{i\Omega(t_2-t_1)})^2(\cosh r-e^{i\Omega
     (t_2-t_1)}\sinh r)^2,\\
     &\gamma_2\gamma_1^*-\gamma_2^*\gamma_1=(e^{i\Omega(t_2-t_1)}-e^{-i\Omega(t_2-t_1)})\gamma^2.
\end{align}
The time dependence of $q_{s_1s_2}$ was determined as a function of the time difference $t_{21}=t_2-t_1$.
\begin{figure}[t]
  \centering
  \includegraphics[width=0.85\linewidth]{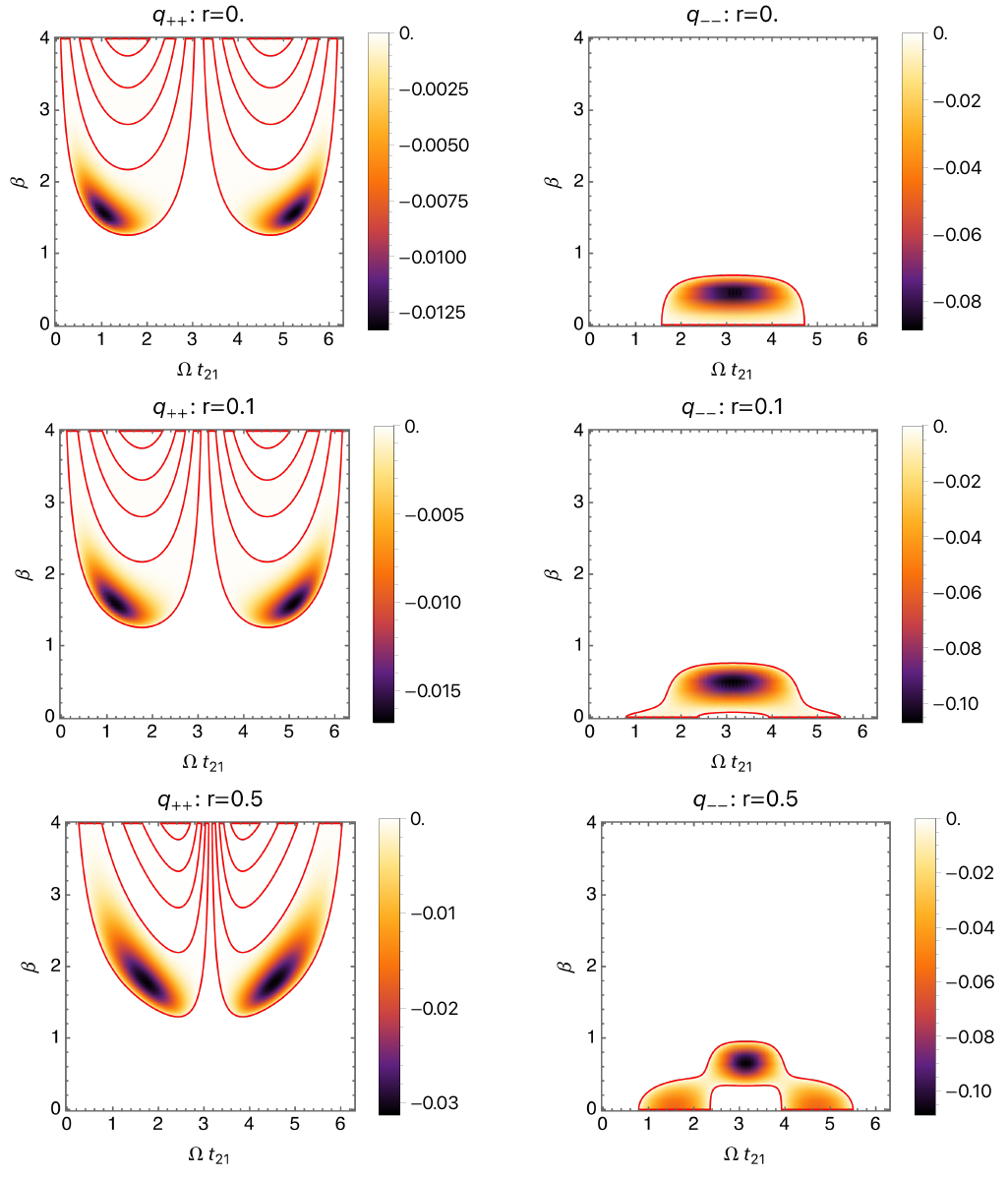}
    \caption{Negative regions of the quasi-probability (enclosed by red lines) for the pure ground state $\ket{0}$ with the 
      squeezed coherent-state projector  $r>0$. $q_{++}$ and
      $q_{--}$ have negative values and the LGIs are
      violated.
    \label{fig:q-sc1}}
 \end{figure}

\begin{figure}[t]
  \centering
  \includegraphics[width=1.\linewidth]{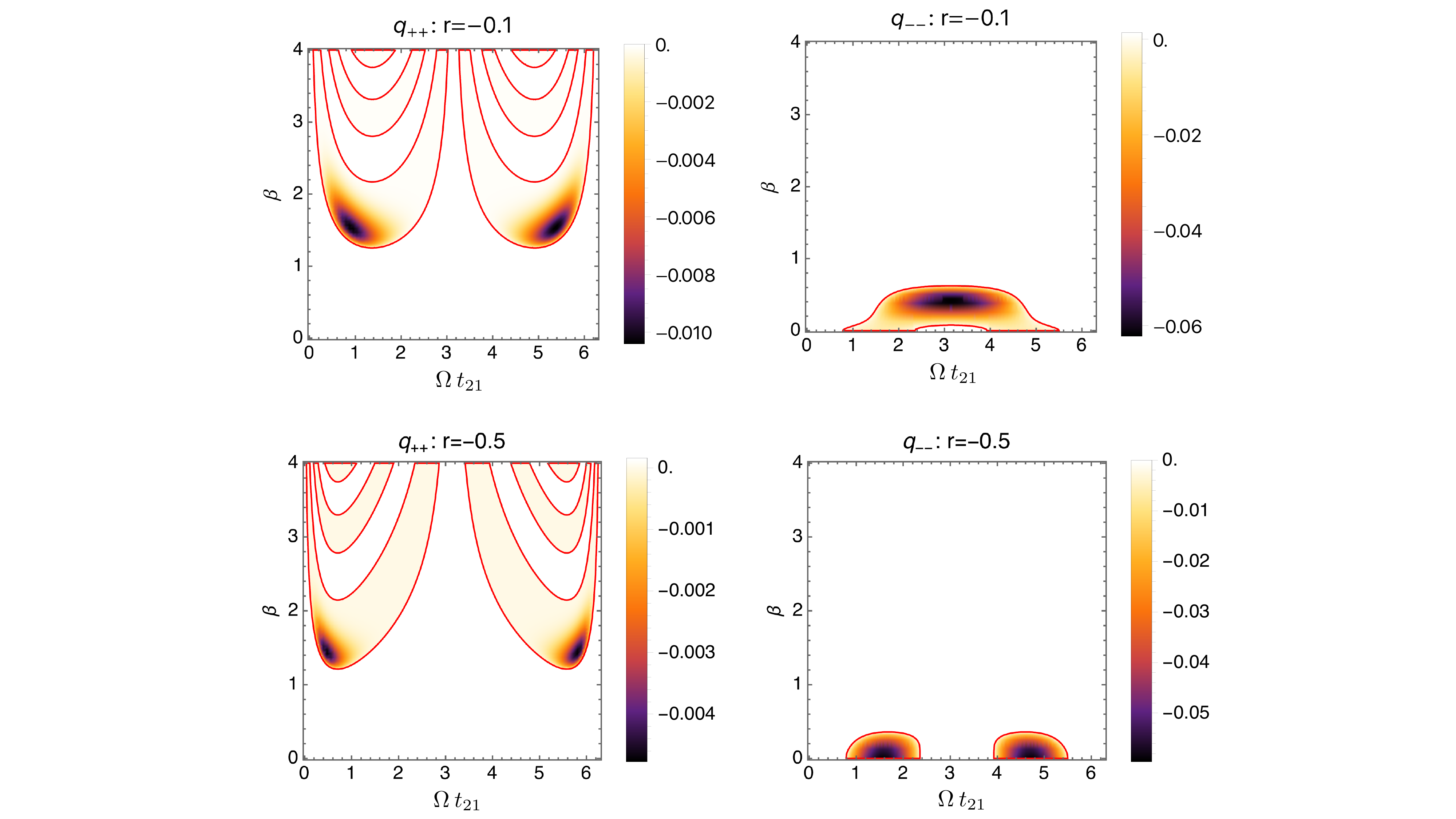}
    \caption{Negative regions of the quasi-probability (enclosed by red lines) for the pure ground state $\ket{0}$ with
      squeezed coherent-state projector  $r<0$. $q_{++}$ and
      $q_{--}$ have negative values and the LGIs are violated.
    \label{fig:q-sc2}}
\end{figure}

Figures \ref{fig:q-sc1} and \ref{fig:q-sc2} show the negative regions of the quasi-probability 
for the vacuum state with a squeezed coherent-state projector. The structures of the negative regions in $q_{++}$ and $q_{--}$ are essentially identical to those in the coherent-state-projector case. However, the squeezing effect of the projector affects the shapes of the negative regions of the quasi-probability and their negative values.

\begin{figure}[t]
  \centering
  \includegraphics[width=0.4\linewidth]{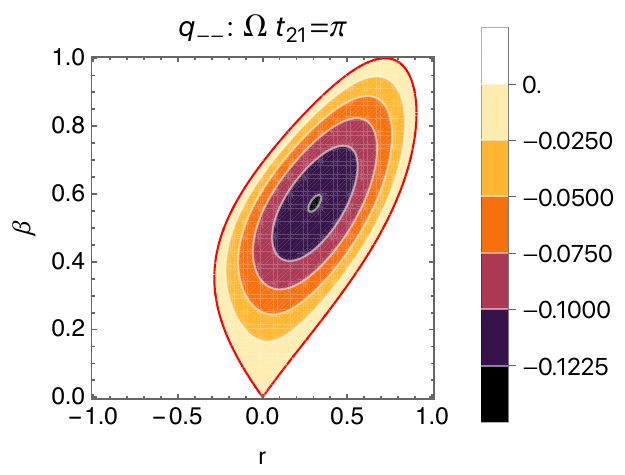}
    \caption{Dependence of $(r,\beta)$ for negative values of $q_{--}$ at $\Omega\,t_{21}=\pi$. At $(r,\beta)\sim (0.31, 0.57)$, the maximum violation of the LGIs close to the L\"{u}ders bound is realized.
    \label{fig:qmini}}
\vspace{1cm}
  \centering
  \includegraphics[width=0.7\linewidth]{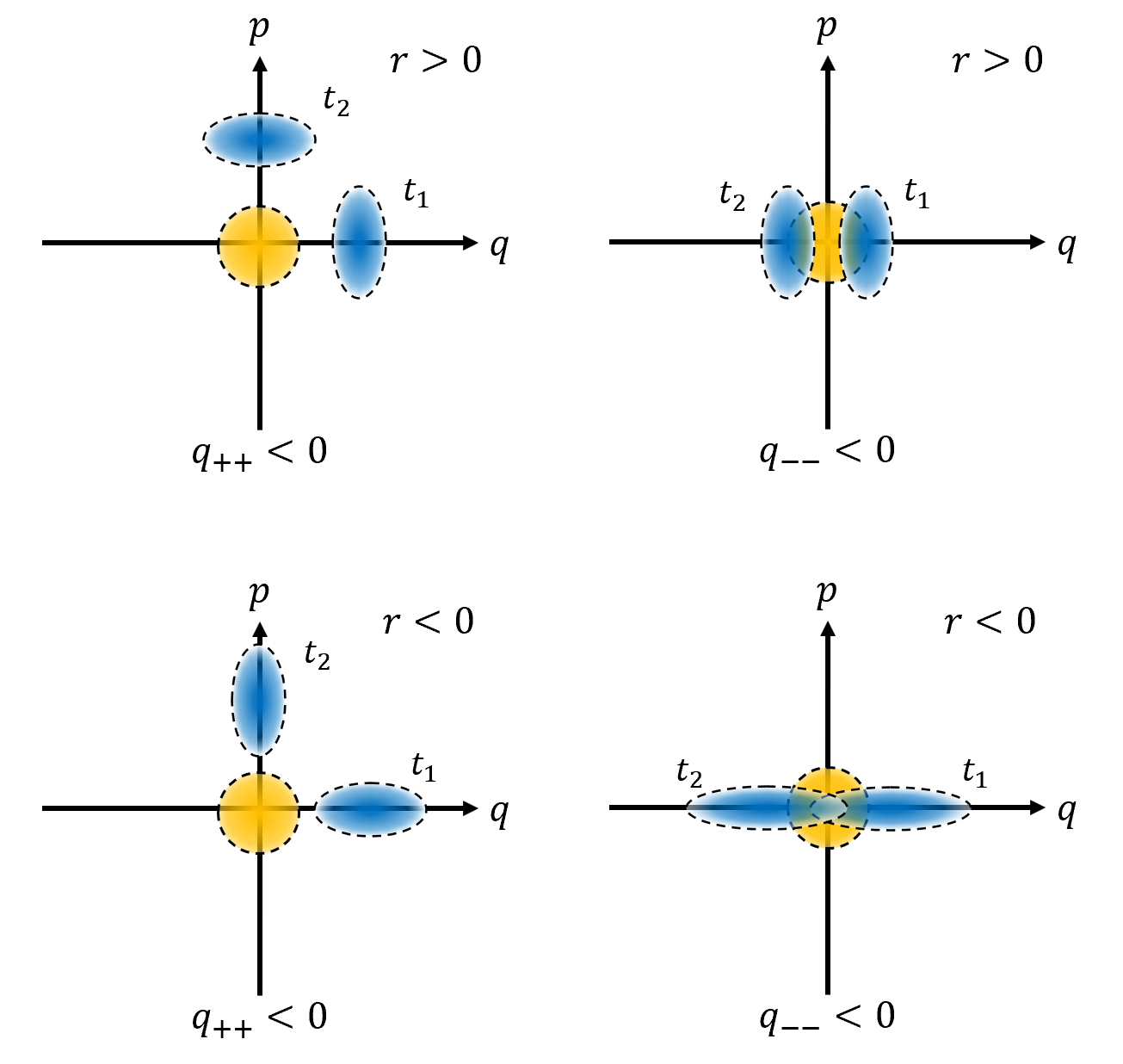}
    \caption{Schematic explanation of the LGIs violation with the squeezed coherent-state projector. The distribution of particles and projectors when the squeezing parameter $r$ is positive is shown in the upper panel, and that when $r$ is negative is shown in the lower panel. $q_{++}$ is the left panel, and $q_{--}$ is the right panel. The deviation of the projector (blue region) from the origin is the coherent amplitude $\beta$. Therefore, we can see that, in $q_{--}$, LGIs violation is occurring between $\beta = 0$ and $\beta = 1$ even when the projector is squeezed.
    \label{fig:q-scm}}
\end{figure}

 First, we consider the $r>0$ case (Figs. \ref{fig:q-sc1} and \ref{fig:qmini}).  We observe that squeezing with $r>0$ enhances the violation of the LGIs in $q_{--}$ (see  Fig. \ref{fig:qmini}, which depicts $(r,\beta)$ dependence of $q_{--}$ at $\Omega\, t_{21}=\pi$). In fact, when $r\approx0.31$, 
    $q_{--}$ can reach $\sim$ $-0.123$ at  $\beta\approx 0.57, \Omega\, t_{21}=\pi$, which is  close to the value of the
    L\"{u}ders bound $-1/8=-0.125$, and a nearly maximal violation of the LGIs is realized. The maximum violation setting of the LGIs is similar to that for  $q_{--}$ with a coherent-state projector for the initial vacuum state  (see $q_{--}$ $(\alpha = 0)$ in Fig. \ref{fig:LGI2}) but,  in terms of the  amount of violation, it becomes greater with the introduction of the squeezing. This reflects the fact that the squeezed state itself represents a quantum nature compared with the simple coherent state without squeezing.
  For $q_{++}$, its values decrease with the increase in $r$. Therefore, squeezing enhances the violations of the LGIs that appears in $q_{++}$.

\begin{figure}[t]
\hspace{-5mm} 
    \centering
    \includegraphics[width=0.7\linewidth]{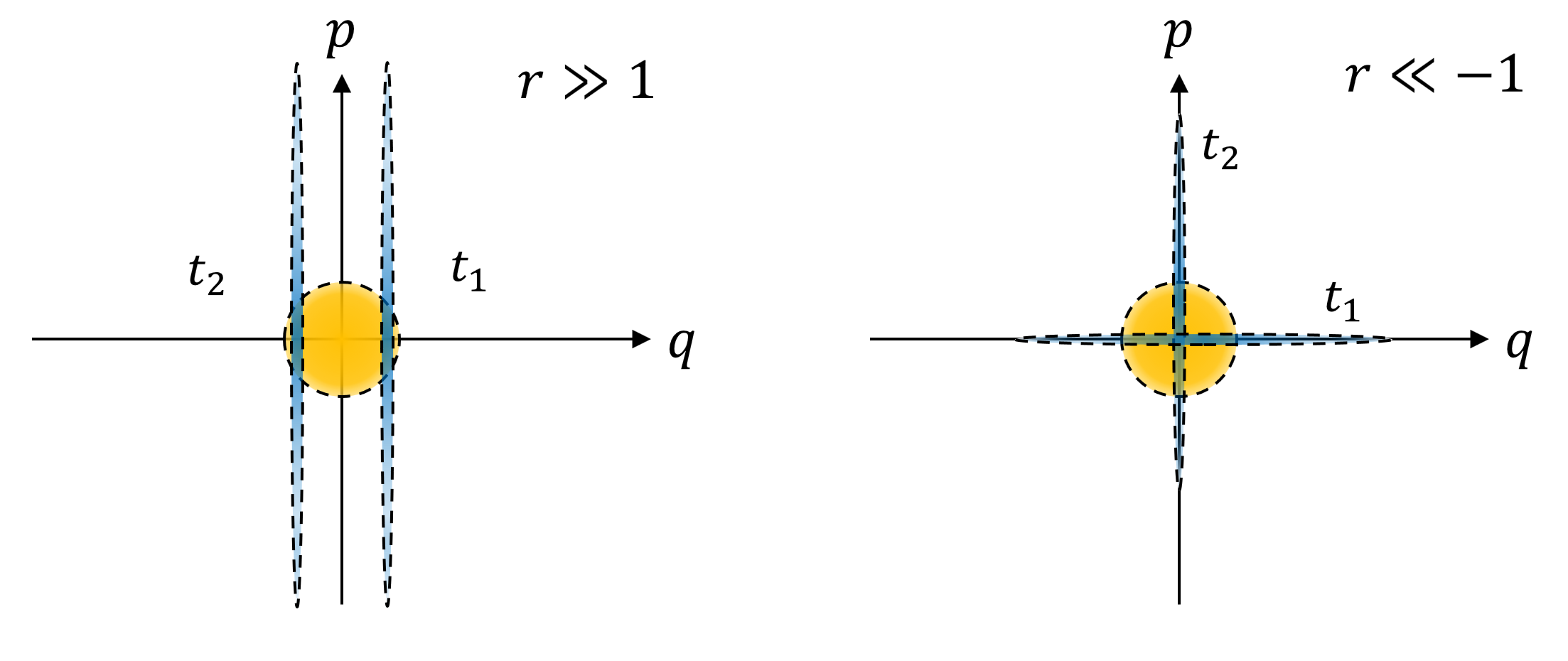}
    \caption{Typical examples of $q_{--}>0$. 
    The left and right panels represent the extreme case of squeezing, in which the LGIs are not violated.
    For large $|r|$, the state after measurement becomes approximately eigenstates of $\hat q$ or $\hat p$, and the violation of the LGIs cannot be expected.}
    \label{fig:qposi}
\end{figure}

Figure \ref{fig:q-scm} shows a schematic explanation of the violation of the LGIs by a squeezed coherent-state projector. This figure helps us understand why quasi-probability is negative in some of the panels of Fig. \ref{fig:q-sc2}. Figure \ref{fig:qposi} conversely illustrates typical examples of a setting in which quasi-probability is not negative (i.e., the LGIs are not violated). The left panel explains why the quasi-probability is not negative in the half period ($\Omega\, t_{21}=\pi$) for the lower right panel in Fig. \ref{fig:q-sc2}. That is, in the orange region, where the particles are predicted to be present, the second measurement measures almost the same region as the first measurement; therefore, we can say that no information is obtained from the second measurement.

 For the $r<0$ case (Figs. \ref{fig:q-sc2} and \ref{fig:qmini}), we observe that the squeezing increases negative values of $q_{++}$ and $q_{--}$ and reduces violation of the LGIs. Therefore, squeezing did not  enhance the violation of the LGIs in this case. As $r$ decreases (e.g., $r=-0.5$), the position of the maximum violation changes from $\Omega\, t_{21}=\pi$ (a half period) to $\Omega\, t_{21}=\pi/2$ (a quarter period) and $\Omega\, t_{21}=3 \pi/2$ (a three-quarter period). As shown in the schematic explanation of the violation of the LGIs (left panel of Fig. \ref{fig:qposi}), this corresponds to the fact that, at $\Omega\, t_{21}=\pi $, we perform the measurement for almost the same area as  measured at $t_1$.  Therefore, we did not obtain any information and the LGIs were not violated.

  For larger absolute values of the squeezing  parameter $|r|\gg 1$, the negative value of the quasi-probability approaches zero for $q_{++}$. This can be explained as follows: A projector with these values of $r$ corresponds to an ideal projective $q$  or $p$ measurement (see the middle and right panels of Fig. \ref{fig:qposi}). 
  These types of measurements project the target state onto the position eigenstate or momentum eigenstate after the measurement. In other words, if $|r|$ is large, the state of the particle collapses to the position or momentum eigenstate, and a violation of the LGIs is not expected.  
  For $q_{--}$, 
  the state is projected to the outside of the blue region, as shown in Fig. \ref{fig:qposi}.
  When $|r|\gg1$, the projected region is almost the same as the original vacuum state, which gives results that are not significantly different from classical motion.  
  This indicates that the information obtained from this measurement was insufficient.   
We observe a similar feature of suppressing the violation of the LGIs for a quantum field with a local projection operator with a large squeezing parameter.


%

%
 %


\section{Violation of LGIs for a quantum field}
As an application of the coherent-state projector, we investigated 
LGIs for a chiral scalar field that appears as an edge excitation of
quantum-Hall systems \cite{Yoshioka2002}. To formulate the LGIs  for the quantum field, we first introduce the spatial
local modes of the quantum field.

\subsection{Chiral scalar field in $(1+1)$-dimensional spacetime}
We consider a quantum chiral scalar field $\hat\varphi$ corresponding to the edge excitation of a quantum Hall system \cite{Hotta2022b,Nambu2023a}. Our main purpose is to investigate the quantum
effect of edge modes in a quantum Hall system, which is
measurable using the local charge density $\pa_{x}\hat\varphi$.
The scalar field obeys the following $(1+1)$-dimensional
Klein--Gordon equation:
\begin{equation}
  \ddot{\hat\varphi}-v^2\,\pa_x^2\hat\varphi=0,
\end{equation}
where $v$ is the propagation speed of edge excitation. It is determined as
\begin{equation}
 v=\frac{c\,U'(y)}{eB}=\frac{cE}{B},
\end{equation}
where $U(y)$ is the trapping potential perpendicular to the edges of the quantum Hall system, $E$ is the electric field induced by $U$, and $B$ is the perpendicular magnetic field. 
Hereafter, we set $v=1$ and adopt $v$ as the units of length and time. 

  The field operator at $x^{+}=t+x$ is expressed
as
\begin{equation}
  \hat\varphi(x^{+})=\int_0^\infty\frac{dk}{\sqrt{4\pi k}}\left[\hat
    a_k\,e^{-ik x^+}+\hat
    a_k^\dag\,e^{ik x^+}\right].
\end{equation}
We assumed a vacuum state in the quantum field by imposing
$\hat a_k\ket{g}=0$.
In our setup, the gauge-invariant physical quantity is the current (charge) density
given by the derivative of the field operator $\hat\varphi$:
\begin{equation}
  \hat\Pi(x^+):=\hat\varphi'(x^+)=-i\int_0^\infty
  dk\sqrt{\frac{k}{4\pi}}
  \left[\hat
    a_k\,e^{-ikx^+}-\hat
    a_k^\dag\,e^{ik x^+}\right].
\end{equation}
The commutators between the field operators are
\begin{align}
  &[\hat\varphi(x^+),\hat\varphi(y^+)]=-\frac{i}{4}\mathrm{sgn}(x^+-y^+),\\
  &[\hat\varphi(x^+),\hat\Pi(y^+)]=\frac{i}{2}\del(x^+-y^+),\\
  &[\hat\Pi(x^+),\hat\Pi(y^+)]=\frac{i}{2}\del'(x^+-y^+).
\end{align}
The two-point correlation functions areas are as follows:
\begin{align}
  &\expval{\{\hat\varphi(x^+),\hat\varphi(y^+)\}}=\frac{1}{2\pi}\int_0^\infty
    \frac{dk}{k}\cos(k(x^+-y^+)),\\
  &\expval{\{\hat\varphi(x^+),\hat\Pi(y^+)\}}=\frac{1}{2\pi}\int_0^\infty
    dk\sin(k(x^+-y^+)),\\
  &\expval{\{\hat\Pi(x^+),\hat\Pi(y^+)\}}=\frac{1}{2\pi}\int_0^\infty
    dk k\cos(k(x^+-y^+)).
\end{align}
The Wightman function for $\hat\varphi$ is
\begin{align}
  D_\varphi(x_1^+,x_2^+)&=\expval{\hat\varphi(x_1^+)\hat\varphi(x_2^+)}=\frac{1}{4\pi}\int_\mu^\infty
                  \frac{dk}{k}e^{-ik(x_1^+-x_2^+-i\ep)}, \notag \\
  &=-\frac{1}{4\pi}\log[\mu(x_1^+-x_2^+-i\ep)],
\end{align}
where we introduce an IR cutoff $\mu$ as the lower bound of the integral
and a UV cutoff $\ep$ by $\ep:=\Delta x$,
 where $\Delta x$ is the spatial cutoff length.
In our analysis,
the chiral scalar field $\hat\varphi$ is an effective one and there exists
a short cutoff length $\ep$, below which the effective treatment of the
edge mode breaks down. In a quantum Hall system, this scale is given
by the magnetic length of the quantum Hall system:
\begin{equation}
  \ell_B=\sqrt{\frac{\hbar}{eB}}.
\end{equation}
We considered the short-length cutoff $\Delta x$ as the length in our
analysis.  The Wightman function for $\hat\Pi$ is as follows:
\begin{align}
  D_\Pi(x_1^+,x_2^+)
  &:=\expval{\hat\Pi(x^+_1)\hat\Pi(x^+_2)}=\pa_{x_1^+}\pa_{x_2^+}D(x_1^+,x_2^+) \notag \\
  &= \frac{1}{4\pi}\int_0^\infty dk k\left[e^{-ik(x_1^+-x_2^+-i\ep)}
    \right]\notag \\
  &=-\frac{1}{4\pi}\frac{1}{(x_1^+-x_2^+-i\ep)^2}.
\end{align}
This quantity exhibits the same behavior as that of the massless scalar
field in four-dimensional spacetime and is independent of the IR cutoff $\mu$.
\begin{figure}[b]
  \centering
  \includegraphics[width=0.8\linewidth]{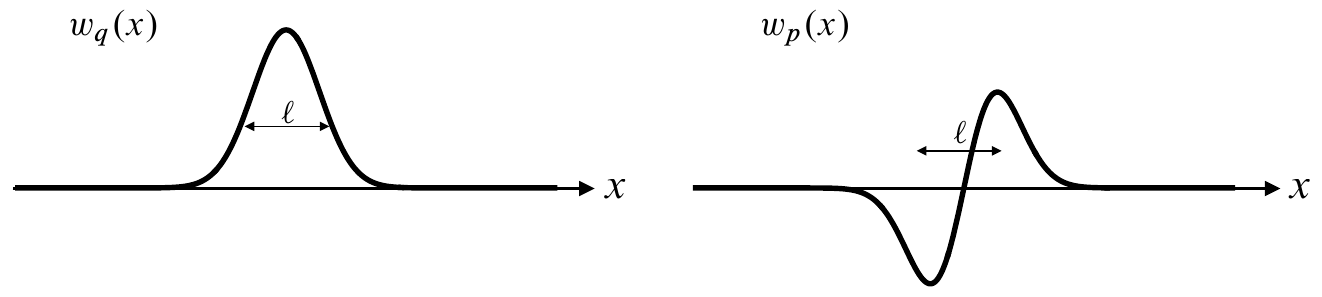}
  \caption{Spatial profiles of window functions $w_q(x)$ and $w_p(x)$.}
    \label{fig:window}
\end{figure}

\subsection{Local spatial  modes of a quantum field}
We considered measurements of the current density
$\hat\Pi(x^+)$ at point $x_A$. The measurement can be represented by the
following interaction Hamiltonian between $\hat\Pi$ and the canonical
variables of the measurement apparatus $(\hat q_D,\hat p_D)$:
\begin{equation}
  \hat H_\text{int}=\lambda(t)g(\hat q_D,\hat p_D)\otimes\int dx\, w_A(x)\,\hat\Pi(t+x),
\end{equation}
where $g(\hat q_D,\hat p_D)$ is a function of the canonical variables of the
measurement apparatus, $w_A(x)$ is a window function that defines the
the spatial mode of the field at $x_{A}$, and $\lambda(t)$ is a
switching function. After acting on the apparatus,
interaction causes a change in the ``reading'' of the apparatus
depending on the state of the quantum field $\hat\Pi$ at $x_A$. In the
present analysis, we did not introduce the details of the measurement protocols
but only focused on the behavior of the local mode of the
quantum field introduced by
the spatial window function.

To measure the values of the field and its conjugate momentum, we define
canonical variables corresponding to the local spatial modes of 
field at $x_{A}$ as follows:
\begin{equation}
  \hat q(t)=\int dx\, w_q(x-x_A)\,\hat\Pi(t+x),\quad   \hat p(t)=\int
  dx\, w_p(x-x_A)\,\hat\Pi(t+x),
  \label{eq:local-mode}
\end{equation}
where $w_{p,q}(x)$ are window functions with nonzero values in a compact
spatial region $x\in[-\ell/2,\ell/2]$.  By requiring these operators to
be canonical pairs, equal-time commutators between these operators
should be
\begin{align}
  &[\hat q, \hat p]=\frac{i}{2}\int dx\, w_q(x-x_A)w_p'(x-x_A)\equiv
    i, 
    \label{eq:s-mode1}\\
  &[\hat q, \hat q]= \frac{i}{2}\int dx\,
    w_q(x-x_A)w_q'(x-x_A)\equiv 0,\\
  &[\hat p, \hat p]=\frac{i}{2}\int dx\,
    w_p(x-x_A)w_p'(x-x_A)\equiv 0.
\end{align}
These conditions are independent of the quantum field state.
Hence, the local spatial  mode $(\hat q,\hat p)$ can be introduced as
suitably choosing  window functions $w_q(x)$ and $w_p(x)$ irrespective of the
states of the quantum field. Locality of the
the spatial mode is guaranteed if window functions are adopted in which
support is compact. We chose the
following $p$-window function in our analysis:
\begin{equation}
w_p(x)=-w_q'(x),\quad \int dx (w_q'(x))^2=2,
\end{equation}
and we assumed that the $q$-window function has the Gaussian form
\begin{equation}
  w_q(x)=2\left(\frac{\ell^2}{\pi}\right)^{1/4}\exp\left(-\frac{x^2}{2\ell^2}\right).
  \label{eq:window}
\end{equation}

The covariances of the local spatial modes are
\begin{align}
  &\langle\hat q^2\rangle=\frac{1}{2}
      \int dx dy\, w_q(x-x_A)w_q(y-x_A)\expval{\left\{\hat\Pi(t+x),\hat\Pi(t+y)\right\}},\\
  &\langle\hat p^2\rangle=\frac{1}{2}
    \int dx dy\,w_p(x-x_A)w_p(x-x_A)\expval{\left\{\hat\Pi(t+x),\hat\Pi(t+y)\right\}},\\
    &\langle\hat q\hat p+\hat p\hat q\rangle=
    \int dx dy\, w_q(x-x_A)w_p(y-x_A)\expval{\left\{\hat\Pi(t+x),\hat\Pi(t+y)\right\}}=0,
\end{align}
where
\begin{equation}
  \expval{\left\{\hat\Pi(t+x),\hat\Pi(t+y)\right\}}=\frac{1}{2\pi}\int_0^\infty
  dk\, k\cos(k(x-y))e^{-\ep k}.
\end{equation}
Therefore,
\begin{align}
  \expval{\hat q^2}&=\frac{\ell^3}{\pi^{3/2}}\int_0^\infty dk\,
        e^{-\ep k}k\left(\int_{-\infty}^\infty dz e^{-z^2/2}\cos(k\ell z)\right)^2
        =\frac{\ell}{\pi^{1/2}}\left(1-\frac{\pi^{1/2}}{2}\left(\frac{\ep}{\ell}\right)+O\left(\frac{\ep}{\ell}\right)^2\right),\\
  \expval{\hat p^2}&=\frac{\ell}{\pi^{3/2}}\int_0^\infty dk\,
        e^{-\ep k}k\left(\int_{-\infty}^\infty dz z
        e^{-z^2/2}\sin(k\ell z)\right)^2=\frac{1}{\pi^{1/2}\ell}\left(1-\frac{3\pi^{1/2}}{4}\left(\frac{\ep}{\ell}\right)+O\left(\frac{\ep}{\ell}\right)^2\right).
\end{align}
  As
$\nu^2=\expval{\hat q^2}\expval{\hat p^2}=1/\pi-5/(4\pi^{1/2})(\ep/\ell)+O(\ep/\ell)^2$, the
state of the local mode is generally thermal (for a pure state, $\nu=1/2$). The mixedness of local
mode depends on the region size $\ell$ and  reflects the entanglement behavior between the local region and its complement. With an increase in $\ell$,  the symplectic eigenvalue $\nu$ increases, and the entanglement between the local modes and their complement becomes greater. This behavior of entanglement of the local mode was discussed in our previous study on a chiral scalar field in a quantum Hall system \cite{Nambu2023a}.



\subsection{Measurement of a quantum field}
We considered coherent-state measurements of the quantum field. The
target local mode  at $x=x_A$ defined by the quantum field is given as
\eqref{eq:local-mode}. The local modes are expressed as follows:
\begin{align}
  &\hat q(t)=-\frac{i}{\sqrt{2}}\int_0^\infty
    dk\, k^{1/2}w_q(k)\left(e^{-ik(t+x_A)}\hat a_k-e^{ik(t+x_A)}\hat
    a_k^\dag\right),\\
  &\hat p(t)=-\frac{1}{\sqrt{2}}\int_0^\infty
    dk\, k^{3/2}w_q(k)\left(e^{-ik(t+x_A)}\hat a_k+
e^{ik(t+x_A)}\hat
    a_k^\dag\right),
\end{align}
where the Fourier component of the window function is introduced as
\begin{equation}
  w_q(k)=\frac{1}{\sqrt{2\pi}}\int_{-\infty}^\infty dx
  \,w_q(x)e^{ikx},\quad w_q(k)=w_q^*(k),\quad w_p(k)=ik w_q(k)
\end{equation}
and, for the window function \eqref{eq:window},
\begin{equation}
  w_q(k)=2\left(\frac{\ell^6}{\pi}\right)^{1/4}e^{-\frac{\ell^2}{2}k^2}.
  \label{windowfield}
\end{equation}
The annihilation operator for the local mode is introduced as follows:
\begin{align}
  \hat b(t):
  &=\sqrt{\frac{\omega}{2}}\,\hat
              q(t)+\frac{i}{\sqrt{2\omega}}\hat p(t)  \notag \\
  &=-\frac{i}{2}\int_0^\infty dk\, k\, w_q(k)\left[
    \left(\sqrt{\frac{\omega}{k}}+\sqrt{\frac{k}{\omega}}\right)e^{-ik(t+x_A)}\hat
    a_k
    -\left(\sqrt{\frac{\omega}{k}}
    -\sqrt{\frac{k}{\omega}}\right)e^{+ik(t+x_A)}\hat
    a_k^\dag\right] \notag \\
  &=-i\int_0^\infty dk\, k\,w_q(k)(\cosh r_k\hat a_k(t)-\sinh r_k\hat a_k^\dag(t)), 
\end{align}
where $\omega$ is a parameter characterizing  the local mode (a parameter characterizing the projector related to the protocols of the measurement apparatus) and  we
introduced $r_k$ and $\hat a_k(t)$ as
\begin{equation}
  e^{r_k}=\sqrt{\frac{\omega}{k}},\quad \hat a_k(t)=e^{-ik(t+x_A)}\hat
  a_k.
  \label{eq:field_squeeze}
\end{equation}
We consider the following Gaussian projector for the local mode at time $t$:
\begin{align}
  &\hat M_s(t)=\frac{1-s}{2}+s\,\hat D_{b(t)}(\beta)\,\hat \rho_{b(t)}\hat
    D_{b(t)}^\dag(\beta), \label{eq:gaussian-measurement}\\
  &\hat
    D_{b(t)}(\beta)=e^{\beta\hat b^\dag(t)-\beta^*\hat b(t)},\quad
  \hat\rho_{b(t)}=\ket{0_{b(t)}}\bra{0_{b(t)}},
\end{align}
where $s=\pm 1$ and $\hat\rho_{b(t)}$ represents the state of
measurement apparatus for the local mode. 
We use the following
formula that connects the displacement operator to the vacuum
density operator:
\begin{equation} \ket{0_{b(t)}}\bra{0_{b(t)}}=\int\frac{d^2z}{\pi}\exp\left[-\frac{1}{2}
    |z|^2\right]\hat D_{b(t)}(z).
\end{equation}
The Gaussian projector can then be written as\footnote{The formulas for
 the displacement operators are
$$ \hat D(\lambda_1)\hat D(\lambda_2)=\hat
  D(\lambda_1+\lambda_2)\exp\{\frac{1}{2}(\lambda_1\lambda_2^*-\lambda_1^*\lambda_2)],\quad
  \hat D(\lambda)\hat D(z)\hat D^\dag(\lambda)=\hat D(z)\exp(-z\lambda^*+z^*\lambda),
$$}
\begin{align}
  \hat M_s(t)&=\frac{1-s}{2}+\frac{s}{\pi}\int d^2z\exp\left[-\frac{1}{2}|z|^2\right]\hat D_{b(t)}(\beta)\hat D_{b(t)}(z)\hat
             D_{b(t)}^\dag(\beta) \notag \\
  &=\frac{1-s}{2}+\frac{s}{\pi}\int d^2z\exp\left[-\frac{1}{2}|z|^2\right]e^{-\beta^*z+\beta z^*}\hat D_{b(t)}(z).
  \label{eq:Gprojector}
\end{align}
We simplify this expression as follows: By using
\begin{align}
  z\hat b^\dag(t)-z^*\hat b(t)
  &=i\int dk\,k\,w_q(k)[(z^*\cosh r_k-z\sinh
    r_k)\hat a_k(t)+(z\cosh r_k-z^*\sinh r_k)\hat a_k^\dag(t)]\notag \\
  &=\int dk\left[Z_k(z,t)\hat a^\dag_k-Z^*_k(z,t)\hat a_k\right], 
\end{align}
with
\begin{equation}
   Z_k(z,t):=ik\,w_q(k)(z\cosh r_k-z^*\sinh r_k)\,e^{ik(t+x_A)},
\end{equation}
the displacement operator for the local mode $\hat D_{b(t)}(z)=e^{z\hat
  b^\dag(t)-z^*\hat b(t)}$ can be written as a product of the displacement operator for each $k$ mode of the original field operator:
\begin{align}
\hat D_{b(t)}(z)&=\prod_k\exp\left[Z_k(z,t)\hat
    a_k^\dag-Z^*_k(z,t)\hat a_k\right]  \notag \\
  &=\prod_k\hat D_{a_k}(Z_k(z,t)).
  \label{eq:Dba}
\end{align}
Therefore, the projector \eqref{eq:Gprojector} can be  expressed by the displacement operator for each $k$ mode of the original field operator.
\subsection{Quasi-probability}
Let us consider the quasi-probability with the initial vacuum state
$\hat\rho_0=\ket{g}\bra{g}$ in a quantum field. The quasi-probability was evaluated as follows:
\begin{align}
  q_{s_1s_2}&=\mathrm{Re}\bra{g}\hat M_{s_2}(t_2)\hat
          M_{s_1}(t_1)\ket{g} \notag \\
  &=\frac{(1-s_1)(1-s_2)}{4}+\frac{(1-s_2)s_1}{2\pi}\,\mathrm{Re}\int
    dz e^{-|z|^2/2-\beta^*z+\beta z^*}\bra{g}\hat D_{b_1}(z)\ket{g}
    \notag \\
  &\quad+\frac{(1-s_1)s_2}{2\pi}\,\mathrm{Re}\int
    dz\, e^{-|z|^2/2-\beta^*z+\beta z^*}\bra{g}\hat D_{b_2}(z)\ket{g}\notag \\
            &\quad+\frac{s_1s_2}{\pi^2}\,\mathrm{Re}\int
          d^2z_2\,d^2z_1e^{-|z_2|^2/2-\beta^*z_2+\beta
          z_2^*}\,e^{-|z_1|^2/2-\beta^*z_1+\beta z_1^*}
          \bra{g}\hat D_{b_2}(z_2)\hat D_{b_1}(z_1)\ket{g}\notag
  \\
        &=\frac{(1-s_1)(1-s_2)}{4}+\frac{(1-s_2)s_1}{2\pi}\mathrm{Re}[I_1]+
          \frac{(1-s_1)s_2}{2\pi}\mathrm{Re}[I_2]+\frac{s_1s_2}{\pi^2}
          \mathrm{Re}[I_3], \label{eq:qs1s2}
\end{align}
where integrals $I_1,$ $I_2,$ and $I_3$ are defined as follows:
\begin{align}
  I_{1,2}&=\int
    d^2z\, e^{-|z|^2/2-\beta^*z+\beta z^*}\bra{g}\hat
           D_{b_{1,2}}(z)\ket{g},\\
  I_3&=\int
          d^2z_2\,d^2z_1\,e^{-|z_2|^2/2-\beta^*z_2+\beta
          z_2^*}\,e^{-|z_1|^2/2-\beta^*z_1+\beta z_1^*}
          \bra{g}\hat D_{b_2}(z_2)\hat D_{b_1}(z_1)\ket{g}.
\end{align}
By using the relation \eqref{eq:Dba}, the expectation values in integrals $I_1,$ $I_2,$ and $I_3$ are evaluated as
\begin{align}
  \bra{g}\hat
  D_{b(t)}(z)\ket{g}&=\prod_k\bra{g}\hat D_{a_k}[Z_k(z,t)]\ket{g}=\prod_k\exp\left[-\frac{1}{2}|Z_k(z,t)|^2\right]
                      \notag \\
  &=\exp\left[-\frac{1}{2}\int dk\, k^2w_q^2(k)|z\cosh r_k-z^*\sinh r_k|^2\right] \notag\\
  &=\exp\left[-\frac{x^2+\ell^2\omega^2y^2}{\pi^{1/2}\ell\omega}\right], \\
  & \notag \\
  \bra{g}\hat D_{b_2}(z_2)\hat D_{b_1}(z_1)\ket{g}
  &=\prod_{k}\bra{g}\hat D_{a_k}[Z_k(z_2,t_2)]\hat
    D_{a_k}[Z_k(z_1,t_1)]\ket{g} \notag \\
  &=\prod_{k}\exp\left[\frac{1}{2}\Bigl(Z_k(z_2,t_2)Z^*_k
    (z_1,t_1)
    -Z^*_k(z_2,t_2)Z_k(z_1,t_1)\Bigr)\right]\bra{g}\hat
    D_{a_k}[Z_k(z_2,t_2)+Z_k(z_1,t_1)]\ket{g}\notag \\
  &=\prod_{k}e^{\frac{1}{2}(Z_2Z_1^*-Z_2^*Z_1)}e^{-\frac{1}{2}(Z_1+Z_2)(Z_1+Z_2)^*}
    \notag \\
  %
  &=\prod_k\exp\Biggl[-k^2w_q^2(k) \biggl(\frac{|z_1\cosh r_k-z_1^*\sinh r_k|^2}{2}+\frac{|z_2\cosh r_k-z_2^*\sinh r_k|^2}{2} \notag \\
   &\qquad\qquad\qquad\qquad
   +e^{-ikt_{21}}(z_2^*\cosh r_k-z_2\sinh r_k)(z_1\cosh r_k-z_1^*\sinh r_k)\biggr)\Biggr] \notag \\
  &=\exp\Bigl[-\frac{2\ell^3}{\pi^{1/2}\omega}\int dk
    \,k\,e^{-\ell^2k^2}
[k^2(x_1^2+x_2^2)+\omega^2(y_1^2+y_2^2)
\notag \\
&
\qquad\qquad\qquad\qquad+2e^{-ikt_{21}}(k
    x_1+i \omega y_1)(k x_2-i\omega y_2)]\Bigr] \notag \\
  &=:\exp(-I_4),
\end{align}
where  $ z_{1,2}=x_{1,2}+iy_{1,2}$ and  $t_{21}=t_2-t_1$.  The integral $I_4$ is
\begin{align}
  I_4&=\frac{1}{4\ell^4\omega}\Biggl[\frac{\ell}{\pi^{1/2}}\left\{-2t_{21}^2x_1x_2+4\ell^4(y_1+y_2)^2\omega^2
    +4\ell^2((x_1+x_2)^2+t_{21}(x_1y_2-x_2y_1)\omega)\right\} \notag \\
  &\qquad\qquad -e^{-\frac{t_{21}^2}{4\ell^2}}\left\{-t_{21}^3x_1x_2+2\ell^2t_{21}(3x_1x_2+t_{21}x_2y_1\omega-t_{21}x_1y_2\omega)+4\ell^4\omega(-x_2y_1+y_2x_1+t_{21}y_1y_2\omega)\right\}
  \notag \\
  &\qquad\qquad\times\left(i+\mathrm{Erfi}\left[\frac{t_{21}}{2\ell}\right]\right)\Biggr].
\end{align}
Therefore, $I_1,$ $I_2,$ and $I_3$ become
\begin{align}
 &I_{1,2}=
    \frac{\pi}{\sqrt{1/2+\frac{1}{\pi^{1/2}\ell\omega}}\sqrt{1/2+\frac{\ell\omega}{\pi^{1/2}}}}\exp\left(-\frac{\beta^2}{1/2+\ell\omega/\pi^{1/2}}\right),\\
%
  &I_3=\int
          d^2z_2\,d^2z_1e^{-|z_2|^2/2-\beta^*z_2+\beta
          z_2^*}e^{-|z_1|^2/2-\beta^*z_1+\beta z_1^*}
          \exp(-I_4(z_1,z_2)).
\end{align}
After performing a Gaussian integral with respect to $z_1$ and $z_2$ in $I_3$, we obtain the exact formula for
$q_{s_1s_2}$ (Eq. \eqref{eq:qs1s2}). The resulting expression has a rather
complicated form and we do not present it here. The quasi-probability depends on the parameters through a combination of $\omega\ell$ and $t_{21}/\ell$. 

\begin{figure}[t]
  \centering
\includegraphics[width=0.85\linewidth]{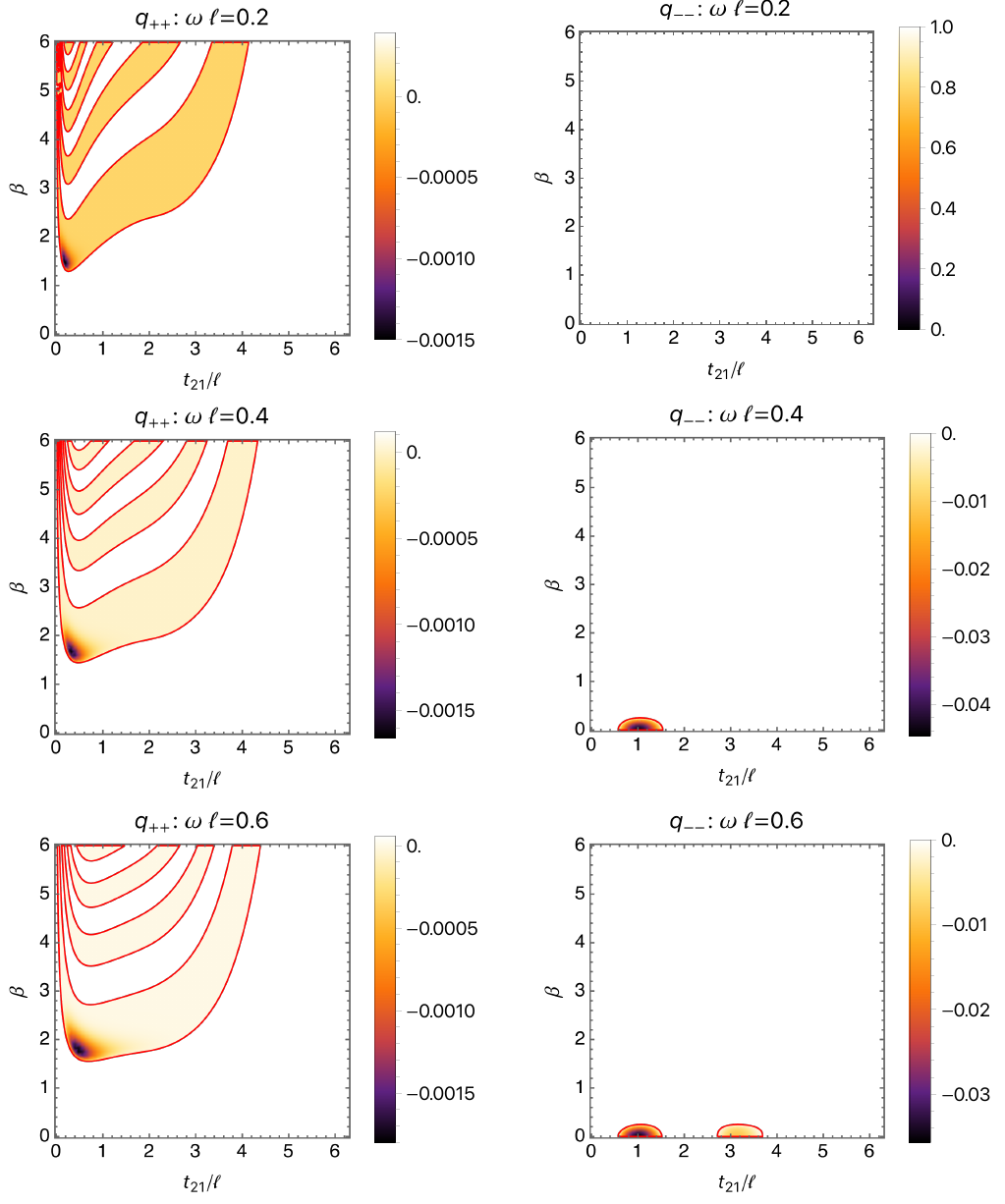}
  \caption{Negative regions of the quasi-probability (enclosed by red lines) for the local mode of the quantum field. The quasi-probabilities $q_{++}$ and $q_{--}$  with $\omega\ell=0.2$, 0.4, and 0.6 are shown.}
  \label{fig:q-field1}
\end{figure}

\begin{figure}[t]
  \centering
\includegraphics[width=0.85\linewidth]{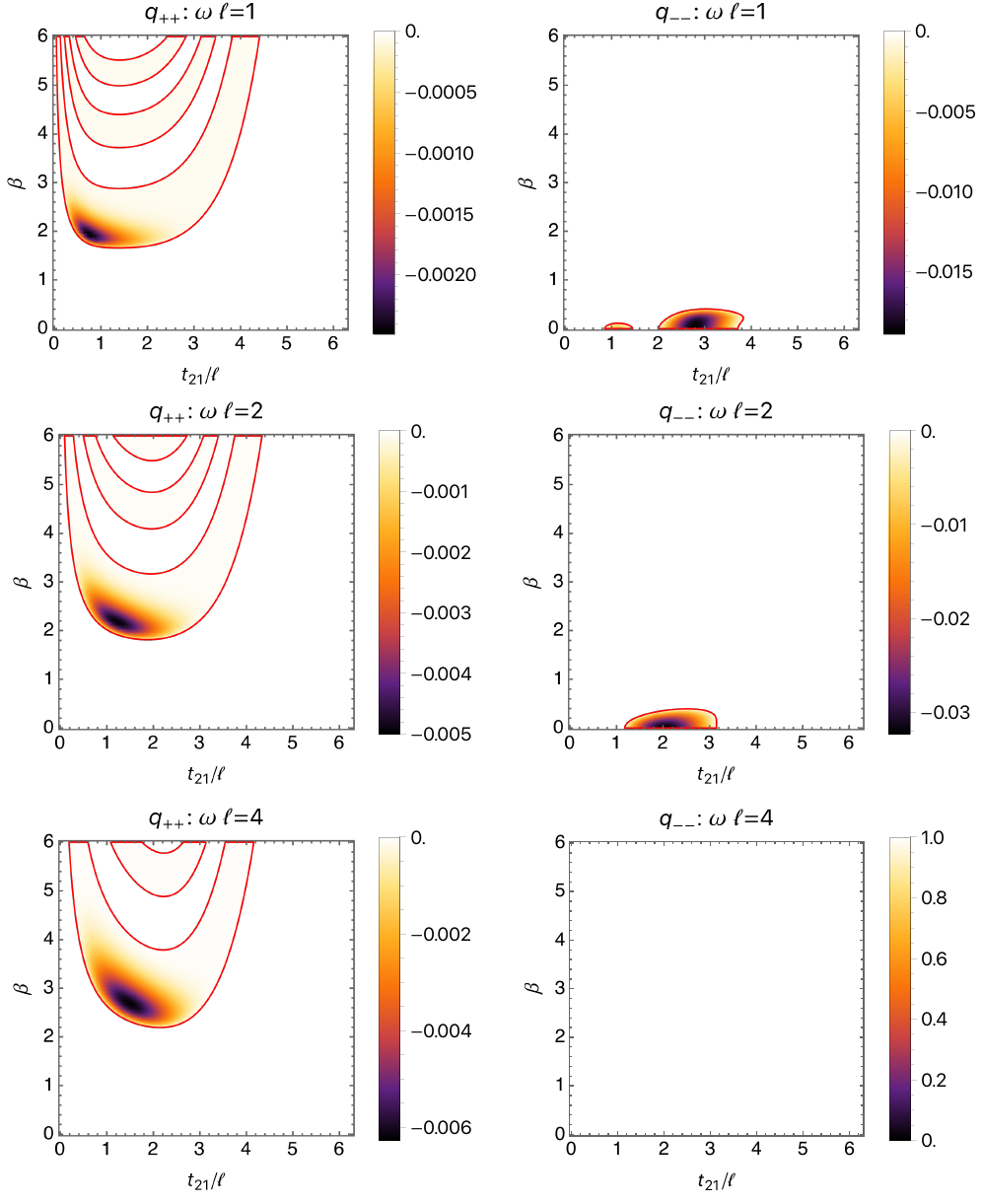}
  \caption{Negative regions of the quasi-probability (enclosed by red lines) for the local mode of the quantum field. The quasi-probabilities $q_{++}$ and $q_{--}$  with $\omega\ell=1$, 2, and 4 are shown.}
  \label{fig:q-field2}
\end{figure}

Figures \ref{fig:q-field1} and \ref{fig:q-field2} show the negative regions of the quasi-probabilities $q_{++}$ and $q_{--}$ in the $(t_{21},\beta)$ plane with values of  $\omega\ell=0.4$--4. 
The structures of the negative regions of $q_{++}$ and $q_{--}$ for the quantum field are similar to those for the harmonic oscillator case, such as the fringe structure of $q_{++}$ and the violation of $q_{--}$ near $\beta \sim 0$. The major difference is the loss of periodicity in  the time direction. Indeed, in $q_{++}$ and $q_{--}$, the symmetry around half of the period with respect to time on the horizontal axis, which is exhibited in the case of harmonic oscillators, is lost. The loss of time periodicity is related to the mixedness of the local mode defined by the quantum field; as the local mode is a subsystem embedded in a total pure system, its state inevitably becomes mixed. This mixedness depends on the size of the local region and the UV cutoff length and reflects the entanglement between the local mode and its complementary degrees of freedom. The evolution of the local mode is nonunitary; Therefore, the periodicity of the state in the time direction is lost. However, the Fourier modes ($k$-mode) are decoupled each other in the Fourier space
and the local mode can be expressed as the sum of the decoupled Fourier 
modes. The loss of periodic features is explained by a dephasing effect.


It is worth mentioning that regions with a negative  $q_{- -}$ disappear when $\omega \ell$ is too large or too small. This behavior can be explained by the fact that $\omega\ell$ corresponds to  the squeezing parameter of the coherent squeezing measurement of the harmonic oscillator. Because the typical wavenumber of the local mode is $k \sim 1/ \ell$, we can regard the parameter $\omega \ell$ as an effective squeezing parameter by comparing \eqref{eq:QHOsqp} and \eqref{eq:field_squeeze}. Hence, when $\omega \ell$ is too large or too small,  too little information is obtained from the measurements (left and right panels in Fig. \ref{fig:qposi}). Therefore, the quantum nature of the measurement is not visible and the LGIs are not violated. For $q_{++}$, the negative value of the quasi-probability approaches zero as squeezing ($\sim$$ \omega \ell$) increases, which is similar to the harmonic oscillator in the previous section.
However, the difference is that the negative regions shift toward larger positive values of $\beta$ as squeezing increases
significantly, although we did not explicitly demonstrate this.

\section{Summary and conclusion}

  We showed violations of the LGIs in terms of the two-time quasi-probability using Gaussian (squeezed) coherent projectors for a harmonic oscillator and a chiral scalar field. For the harmonic oscillator, violations appear mainly in $q_{++}$ and $q_{--}$ even in vacuum. In the thermal state, the LGIs are not violated when the temperature is sufficiently high. When the projector is prepared in a squeezed coherent state, the quasi-probability reaches $-0.123$ for $r\sim 0.31$ and $\beta \sim 0.57$, which is equivalent to $98\%$ of the L\"{u}ders bound.
  Generally, a violation of LGIs occurs when the measurement results differ from those classically expected.
  Therefore, the violation of LGIs represents a feature of the quantumness of the systems; however, further study is required to clarify why a violation close to the L\"{u}ders bound is obtained.

We also showed violations of the LGIs  for a local mode in the chiral scalar field, similar to those for the harmonic oscillator. A major difference is the disappearance of the periodicity in violations. The reason for this periodicity loss is that the local mode is in a mixed state as a subsystem of the entire system. 
This is consistent with previous results ~\cite{2023arXiv231102867T}.
This could be related to the entanglement entropy of the local mode. 
The violations of the LGI observed here are more significant than those
with the simple projection operators reported in Tani \textit{et al.}~\cite{2023arXiv231102867T}. 


Our results can be applied to experiments on LGIs using a quantum Hall edge system \cite{Hotta2022b}. 
Because we use the Gaussian POVM measurement of local modes, we should obtain a dichotomic value of $\hat Q(t)=s(2 \hat M_s(t)-1)$  by measuring $(q,p)$, 
where $\hat{M}_s$ is given by Eq.~(\ref{eq:gaussian-measurement}).
To this end, we consider adjacent regions A and B and measure their local values $q_A$ and $q_B$ using the window function $w_q(x)$:
$\hat q_A=\int dx\, w_q(x-x_A)\,\hat \Pi(t+x)$
and 
$\hat q_B=\int dx\, w_q(x-x_A-\ell)\,\hat \Pi(t+x)=\int dx\, w_q(x-x_A)\,\hat\Pi(t+x+\ell)$, where the window function 
is defined by (\ref{windowfield}).
Using $\hat q_A$ and $\hat q_B$, which include the information of the momentum because their difference 
is equivalent to the momentum essentially, 
makes it possible to obtain $Q$ to perform a coherent measurement \eqref{eq:gaussian-measurement}.

 As well as an application of our theoretical predictions
to quantum Hall experiments, applying them to harmonic oscillator systems would also be interesting. Recently, macroscopic oscillators have attracted 
attention as a means to explore the boundary between the worlds of quantum and classical mechanics
(e.g., \cite{Westphal:2020okx,Whittle:2021mtt,2020arXiv200810848S}). 
In such macroscopic oscillators, techniques such as continuous measurements, feedback control, and optimal filters are used to generate quantum states. There is also the issue of constructing dichotomic measurements
in macroscopic oscillators. Therefore, applying our predictions to such systems is not trivial and further investigation is necessary.

\begin{acknowledgements}
  We thank Masahiro Hotta for providing valuable insight into this subject. 
Y.N. was supported by JSPS KAKENHI (Grant Nos. JP22H05257 and JP19K03866).
A.M. was supported by JSPS KAKENHI (Grant Nos. JP23K13103 and JP23H01175).
K.Y. was supported by JSPS KAKENHI (Grant Nos. JP22H05263 and JP23H01175). 

\end{acknowledgements}

\appendix

\section{Another derivation of $q_{s_1s_2}$ with a squeezed coherent-state projector}
We can derive \eqref{eq:qsc} without using the properties of the squeezed coherent state. The initial vacuum state is considered as follows: 
\begin{equation}
  \hat\rho_0= \ket{0_a}\bra{0_a}.
\end{equation}
The measurement operator with the ``vacuum'' seed state is
\begin{align}
  \hat M_s(t)&=\frac{1-s}{2}+s\hat
               D_{b(t)}(\beta)\ket{0_{b(t)}}\bra{0_{b(t)}}\hat
               D^\dag_{b(t)}(\beta) \notag \\
  &=\frac{1-s}{2}+\frac{s}{\pi}\int d^2z\, e^{-|z|^2/2}\hat
    D_{b(t)}(\beta)\hat D_{b(t)}(z)\hat D_{b(t)}^\dag(\beta) \notag \\
  &=\frac{1-s}{2}+\frac{s}{\pi}\int d^2z\, e^{-|z|^2/2-\beta^*z+\beta
    z^*}\hat D_a[Z(t,z)],
\end{align}
where $Z(t,z)=(z\cosh r-z^*\sinh r)e^{i\Omega t},
e^{2r}=\omega/\Omega$. The quasi-probability is
\begin{align}
  q_{s_1s_2}&=\mathrm{Re}\bra{0_a}\hat M_{s_2}(t_2)\hat M_{s_1}(t_1)\ket{0_a}
              \notag \\
  &=\frac{(1-s_1)(1-s_2)}{4}+\frac{(1-s_2)s_1}{2\pi}\int
    d^2z_1\,
    e^{-|z_1|^2/2-\beta^*z_1+\beta z_1^*}\expval{\hat
    D_a[Z(t_1,z_1)]} \notag \\
  &\quad +\frac{(1-s_1)s_2}{2\pi}\int
    d^2z_2\,
    e^{-|z_2|^2/2-\beta^*z_2+\beta z_2^*}\expval{\hat
    D_a[Z(t_2,z_2)]} \notag \\
  &\quad +\frac{s_1s_2}{\pi^2}\mathrm{Re}\int d^2z_1d^2z_2\,e^{-|z_1|^2/2-\beta^*z_1+\beta z_1^*-|z_2|^2/2-\beta^*z_2+\beta z_2^*}\expval{\hat
    D_a[Z(t_2,z_2)]\hat D_a[Z(t_1,z_1)]}.
    \label{eq:A3}
\end{align}
The expected values of the displacement operators are
\begin{align}
 &\expval{\hat D_a[Z(t_1,z_1)]}=\expval{\hat
   D_a[Z(t_2,z_2)]}=\exp\left(-\frac{1}{2}|z\cosh r-z^*\sinh
   r|^2\right),\\
  &\expval{\hat
    D_a[Z(t_2,z_2)]\hat D_a[Z(t_1,z_1)]}=
    \expval{\hat
    D_a(Z(t_1,z_1)+Z(t_2,z_2))}\exp\left[\frac{1}{2}(Z_2Z_1^*-Z_1Z_2^*)\right]
    \notag \\
  &=\exp\left(-\frac{|Z_1|^2}{2}-\frac{|Z_2|^2}{2}-Z_1Z_2^*\right).
\end{align}
After performing the Gaussian integrals in \eqref{eq:A3}, we obtain $q_{s_1s_2}$ using the
squeezed coherent-state projector.

%

\end{document}